# Review of meta-heuristic optimization algorithms to tune the PID controller parameters for automatic voltage regulator

*Md. Rayid Hasan Mojumder[1,#], Naruttam Kumar Roy[1]*

[1]Department of Electrical and Electronic Engineering, Khulna University of Engineering & Technology, Khulna 9203, Bangladesh

[#]Corresponding author. Email: rayidmojumder@gmail.com

**Abstract:**

A Proportional- Integral- Derivative (PID) controller is required to bring a system back to the stable operating region as soon as possible following a disturbance or discrepancy. For successful operation of the PID controller, it is necessary to design the controller parameters in a manner that will render low optimization complexity, less memory for operation, fast convergence, and should be able to operate dynamically. Recent investigations have postulated many different meta-heuristic algorithms for efficient tuning of PID controller under various system milieus. However, researchers have seldom compared their custom made objective functions with previous investigations while proposing new algorithmic methods. This paper focuses on a detailed study on the research progress, deficiency, accomplishment and future scopes of recently proposed heuristic algorithms to designing and tuning the PID controller parameters for an automatic voltage regulator (AVR) system. Objective functions, including ITSE, ITAE, IAE, ISE, and ZLG, are considered to enumerate a measurable outcome of the algorithms. Considering a slight variation in the sytem gain parameters of the algorithms, the observed PID gain with ITSE results in 0.81918 - 1.9499 for $K_p$, 0.24366 - 1.4608 for $K_i$, and 0.31840 - 0.9683 for $K_d$. Whereas with ITAE the values are 0.24420 - 1.2771, 0.14230 - 0.8471, and 0.04270 - 0.4775, respectively. The time domain and frequency domain characteristics also changes significantly with each objective function. Our outlined comparison will be a guideline for investigating any newer algorithms in the field.

## I. Introduction:

The automatic voltage regulator (AVR) of a synchronous generator (SG) is essentially used in power system utilities to improve the voltage stability of the system. Any drop in the generator terminal voltage can result in an increased line losses, voltage fluctuations, damage to the loads, and causes financial issues. Thus, it is required that the output voltage is efficiently controlled. The AVR governs the terminal voltage of an SG to keep it to the nominal rating under various system phenomena (no-load, full load, half load, disturbance, and others) by altering the exciter voltage of the generator [1]. In a generator, the large inductance value of the field windings and rapid load fluctuations reduces the dynamic response of the AVR [2]. Therefore, improving the dynamic response of the AVR is a must. In this regard, diverse control techniques comprising of Proportional Integral Derivative (PID) [3], Proportional Integral Derivative Acceleration (PIDA) [4], Fraction Orders PID (FOPID) [5], Sugeno Fuzzy Logic (SFL) [6] controllers have been investigated by researchers to address and to improve the AVR system dynamics. However, a PID controller is the most preferable due to its simple design structure and robustness in operation [7].

Though the augmentation of the controller segment can somewhat improve the AVR system dynamics, the controller can not maintain stability during variable operating points, nonlinear loads, and time delays. The drawbacks are addressed by tuning the controller gain parameters in dynamic system presets. In the case of the PID controller, the gain parameter values can be enumerated by considering trial and error [8], conventional Ziegler Nichols (ZN), or Cohen Coon (CC) methods ([9]). The trial-and-error process requires a tremendous amount of time to converge and seldom yields any optimal result. The traditional ZN and CC method adjusts the gain parameters by considering linear system modeling for an operating point. Neither of the two ways can efficiently withstand system non-linearity during discrepancy and often results in undesirable overshoots along with long-term oscillations in the system.

Moreover, both the methods require a higher load of numerical analysis to extract the optimal parameters of the PID controllers and, thus, are inefficient. Optimization techniques are introduced to overcome the problem of PID gain parameters tuning. Intelligent optimization techniques can adjust the system in varying system conditions. Artificial intelligence optimization techniques fall in the category of neural network and fuzzy logic [10]. In an artificial neural network, the training process considers a significant amount of data and higher convergence. In the fuzzy logic system, the system efficiency depends on the analysis of data, tuning of a model, and the designer's competency during the creation of fuzzy membership functions [11]–[13]. Thus, both of the methods tune differently based on the device, amount of available data, computation complexity and mindset of the operator.

In recent times, meta-heuristic optimization algorithms are getting more attention due to their tunability of controller parameters in a much simpler and easier way without requiring any information gradient. Such optimization algorithms can address system fluctuations in a dynamic environment with significant efficacy. The heuristic algorithms can primarily be categorized under four different groups: swarm-based algorithms, physics-based algorithms, human-based algorithms, and evolutionary algorithms. Researchers have introduced different types of swarm-based algorithms that includes Particle Swarm Optimization (PSO) [14], Artificial Bee Colony

(ABC) [15], Bat Search (BAT) [16], Ant Colony Optimization (ACO) [17], Cuckoo Search (CS) [18], Many Optimizing Liaisons (MOL) [19]. Physics-based algorithms group is comprised of Gravitational Search Algorithm (GSA) [20], Bio-geography-Based Optimization (BBO) [21]. Under the human-based algorithms, there are Teaching Learned Based Optimization (TLBO) [22], Harmony Search Algorithm (HSA) [23]. And under evolutionary algorithms, there are Genetic Algorithm (GA) [24], Local Unimodal Sampling (LUS) [25]. Among other algorithms there are Grasshopper Optimization Algorithm (GOA) [26], Chaotic Ant Swarm (CAS) [27], Continuous Firefly Algorithm (CFA) [28], Salp Swarm Algorithm (SSA) [29], Monarch Butterfly Optimization (MBO) [30], Ant Lion Optimizer (ALO) [31], Bacterial Foraging Optimization Algorithm (BFOA) [32], Symbiotic Organisms Search (SOS) algorithm [33], Taguchi Combined Genetic Algorithm (TCGA) [34], Anarchic Society Optimization (ASO) [35], Pattern Search Algorithm (PSA) [36], Differential Evolution (DE) [37], Chaotic Optimization-based on Lozi Map (COLM) [38], Stochastic Fractal Search (SFS) algorithm [39], Sine Cosine Algorithm (SCA) [40], World Cup Optimization (WCO) [41], Improved Kidney-inspired Algorithm (IKA) [42] and Tree Seed Algorithm (TSA) [43]. The corresponding authors have tried to investigate the transient response, root locus, and bode diagrams for the optimal PID-AVR system design using the algorithm.

Moreover, the robustness of the proposed models is depicted by changing the exciter, and generator models transfer function gains as a dummy for system disturbance.

In most cases, the proposed system improves the static and dynamic system responses significantly. During comparing each of the newly proposed algorithms, aforementioned researchers only considered a handful of previous algorithms. Thus, it is required to gather all the significant algorithms to a platform for outlining the research achievement on this PID-based AVR system design as a whole. However, till now, no research attempt has been performed to address the issue.

In this work, a comprehensive analysis of some of these essential heuristic algorithms is performed. This paper presents the core basis of the algorithms, their mathematical modeling, flow charts, system modeling, performance evaluation, and mutual comparison by accumulating the information from the previous literature. This paper focuses on the performance metrics by considering the root-locus plot, bode diagram investigation, and step response in particular. The objective functions that have been used for the evaluation of the proposed algorithm are necessarily considered.

In brief, the contributions of this paper are outlined as:

1) Analysis of the current research trends on the PID-based controlling of AVR.
2) Algorithms to optimize the PID controller gains for static and robust system operation.
3) Introduction and performance evaluation of the meta-heuristic algorithms for PID controller gain adjustments considering different objective functions.
4) Underlining some key factors to be considered during the future investigation of newer control algorithms.

The paper is structured as follows. Section II briefly represents the basic PID controller model and operation. In section III, a generic AVR system is provided with conventional transfer functions and gain parameters. Section IV outlines few principal objective functions for evaluating the algorithms. In section V, a detailed investigation of the algorithms one after another is provided. Then, in section VI, the comparative performance appraisal among the algorithms is demonstrated. In Section VII, the actual improvement, current paucity, and future prospects of the PID-control algorithms in the application of AVR are outlined. Finally, Section VIII concludes the paper. Fig. 1 depicts the organization of the paper.

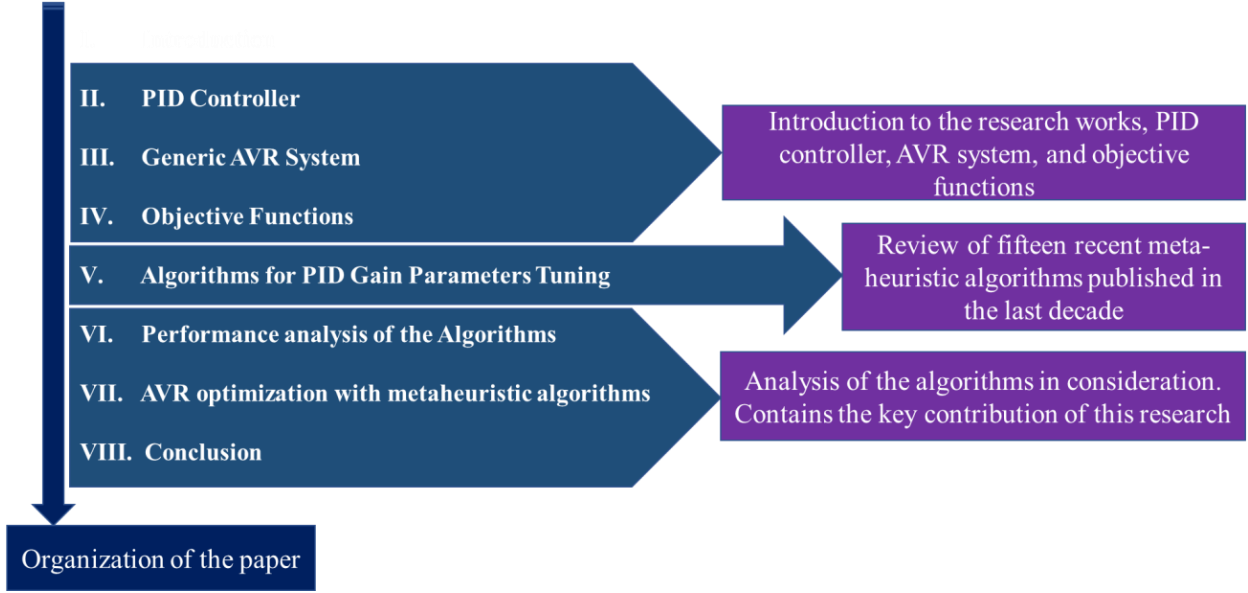

Fig. 1. Organization of this research work.

## II. PID Controller

PID controllers are the most widely used controller in industrial applications. A PID controller is popular due to its simple structure and reliable performance over a wide range of operating conditions. It is used to improving the dynamic response of a system as well as to abate the maximum overshoot, settling time, rise time, and the error at the steady-state condition of operation. A generic PID controller comes with three gain parameters: proportional gain ($K_p$), integral gain ($K_i$), and derivative gain ($K_d$). The $K_p$ controls the rise time performance and is used to tune the rise time performance. The controller gain ($K_i$) adds a pole at the origin, and due to this, the system type improves by one. The $K_i$ also reduces the steady-state error due to a step function to zero. The $K_d$ adds a finite zero to the open-loop plant and reduces the system overshoot and improves the stability margin. The schematic structure of a PID controller is shown in Fig. 2. The transfer function of the PID control system is:

$$G_{PID}(s) = \frac{u(s)}{e(s)} = K_p + \frac{K_i}{s} + K_d s = K_p\left(1 + \frac{1}{T_i s} + T_d s\right), \qquad (1)$$

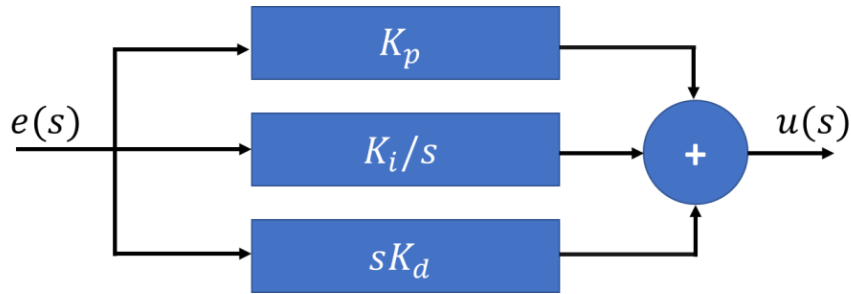

Fig. 2. Schematic diagram of a PID controller.

where the control signal *u(s)* is produced from the error e(s) between the reference and desired signal, the integral action time and derivative action time are presented by $T_i$ and $T_d$, respectively. Prior to putting the controller action, to satisfy control objectives, it is required to collectively tune the three PID gains to obtain a closed-loop system.

### III. Generic AVR System

An Automatic Voltage Regulator (AVR) uses excitation control to enhance the voltage stability of power systems [38]. In a synchronous generator, the role of AVR is the dispatch of primary voltage control to keep the terminal voltage at a nominal value under different system operations. An AVR controls the field exciter to maintain the output voltage in an economical way [44]. A generic AVR system comprises four sub-system structures, as shown in Fig. 3 [39]. In literature, the sub-systems are coined as the amplifier, exciter, generator, and sensor. For control system analysis, the sub-systems' transfer functions are linearized by dropping any saturation or non-linearity. In such a case, the transfer function contains only the significant time constants and gain parameters of the sub-system. Fig. 4 depicts the closed-loop AVR system with four first-order linearized transfer blocks.

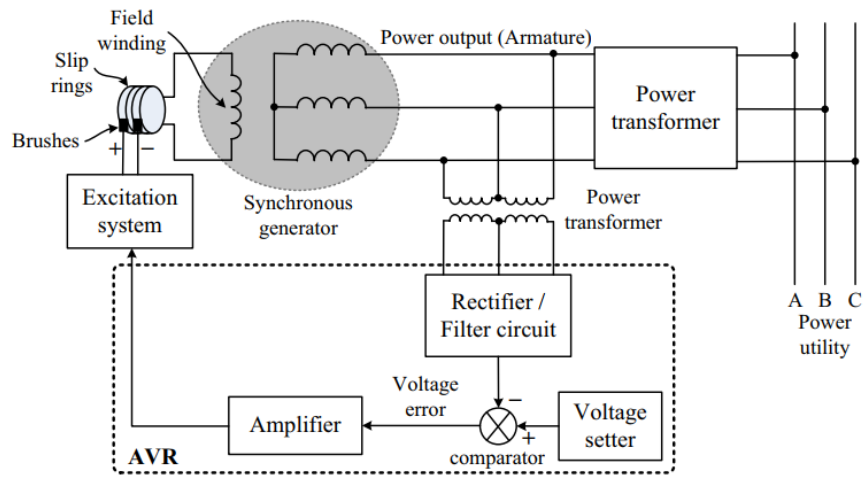

Fig. 3. Schematic diagram of an AVR arrangement [39].

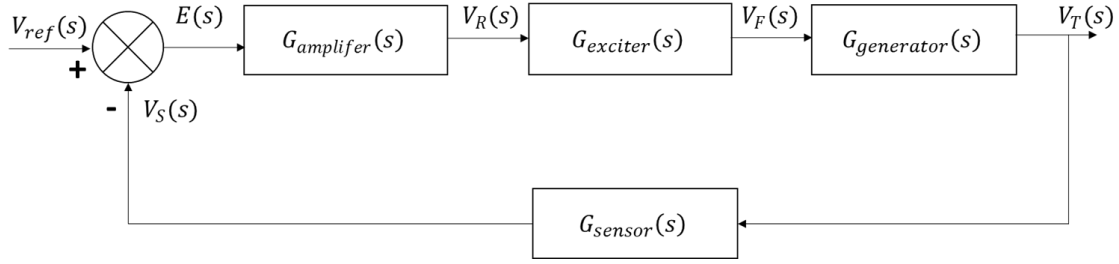

Fig. 4. Block diagram of an AVR system.

Here $\Delta V_{ref}(s)$ is the change in the reference voltage, $\Delta V_s(s)$ is the change in sensor output and $\Delta V_t(s)$ is the change in the terminal voltage value, all in the s-domain. The gain and time constant range for the four sub-system blocks can be observed in Table. 1.

Table. 1. Time constant and gain parameter values of an AVR system [15], [42], [45], [46].

| Components | Ranges of gain parameters | Ranges of time constants (sec) | Chosen parameter values |
|---|---|---|---|
| Amplifier | $10 \leq K_A \leq 40$ | $0.02 \leq T_A \leq 0.1$ | $K_A = 10, T_A = 0.1$ s |
| Exciter | $1.0 \leq K_E \leq 10$ | $0.5 \leq T_E \leq 1.0$ | $K_E = 1.0, T_E = 0.4$ s |
| Generator | $0.7 \leq K_G \leq 1.0$ | $1.0 \leq T_G \leq 2.0$ | $K_G = 1.0, T_G = 1.0$ s |
| Sensor | $1.0 \leq K_S \leq 2.0$ | $0.001 \leq T_S \leq 0.06$ | $K_S = 1.0, T_S = 0.01$ s |

The four sub-system structures of the AVR are modeled as follows:

1) The amplifier model considers an amplifier gain K$_A$ and a time constant T$_A$. Amplifier block converts the error signal E(s) to exciter block input voltage V$_R$(s). Thus, the transfer function is:
$$G_{amplifer}(s) = \frac{V_R(s)}{E(s)} = \frac{K_A}{1+T_A s} \qquad (2)$$

2) The exciter system converts the input voltage V$_R$(s) to the exciter circuit to V$_F$(s) to the field circuit. The exciter model consists of gain K$_E$ and the time constant for the system is T$_A$. The transfer function is:
$$G_{exciter}(s) = \frac{V_F(s)}{V_R(s)} = \frac{K_E}{1+T_E s} \qquad (3)$$

3) In the linearized model, the transfer function relating the generators terminal output voltage V$_T$(s) to the field excitation voltage V$_F$(s) is given by using generator gain K$_G$ and time constant T$_G$:
$$G_{generator}(s) = \frac{V_T(s)}{V_F(s)} = \frac{K_G}{1+T_G s} \qquad (4)$$

The generator time constant T$_G$ is load-dependent.

4) The sensor model senses the generator terminal voltage V$_T$(s) and feeds sensor output voltage V$_S$(s) to the system node where V$_S$(s) is compared with the reference voltage level V$_{Ref}$(s). If the sensor controller gain is K$_G$ and the time constant is T$_A$, the transfer function appears as:
$$G_{sensor}(s) = \frac{V_S(s)}{V_T(s)} = \frac{K_S}{1+T_S s} \qquad (5)$$

In Table. 1, the model's gain and time constant range are addressed. Using the above four model transfer equation, for the AVR system, the closed-loop transfer function is:

$$G_{AVR}(s) = \frac{V_T(s)}{V_{ref}(s)} = \frac{K_A K_E K_G (1+T_S s)}{(1+T_A s)(1+T_E s)(1+T_G s)(1+T_S s) + K_A K_E K_G K_S} \qquad (6)$$

The dynamic behavior of this close loop AVR can be obtained by considering a unit step response. For this, firstly, the gain and time constant are chosen as per Table 1. The Simulink model of the AVR system without any controller and with a step input is shown in Fig. 5 [43]. The terminal voltage response to the step input of the AVR system is shown per unit (p.u.) in Fig. 6 [43]. From this figure, it can be seen that the output voltage does not track the desired reference value (1 p.u.) well. The AVR systems open-loop characteristics maximum overshoot percentage (Mp%), rise time (Tr), settling time (Ts), peak time (Tp), and steady-state error (Ess) are accumulated in Table. 2. In Table. 3, the number of pole-zeros and their positions are given. From tables 1 and 2, it is observed that due to two complex poles near the imaginary axis, the system becomes oscillatory damping and requires high rise time and settling time. Thus, for improving the dynamic response, a controller arrangement is needed.

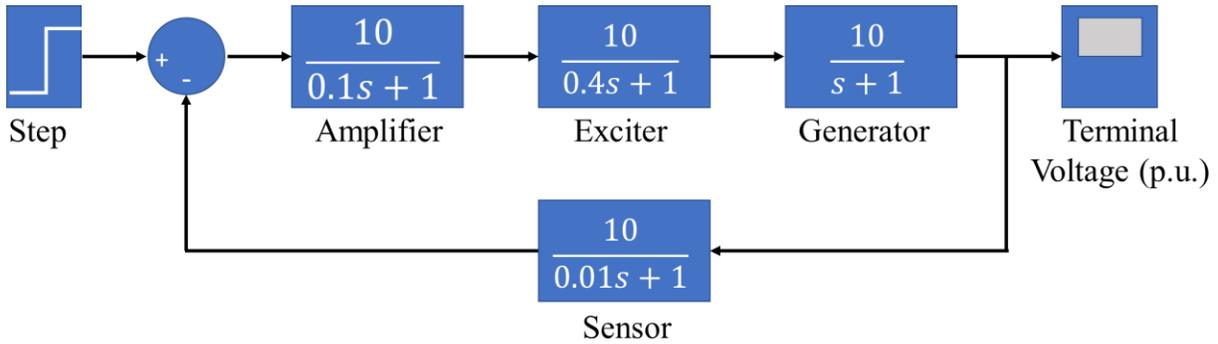

Fig. 5. Simulink modeling of an AVR system without any PID controller.

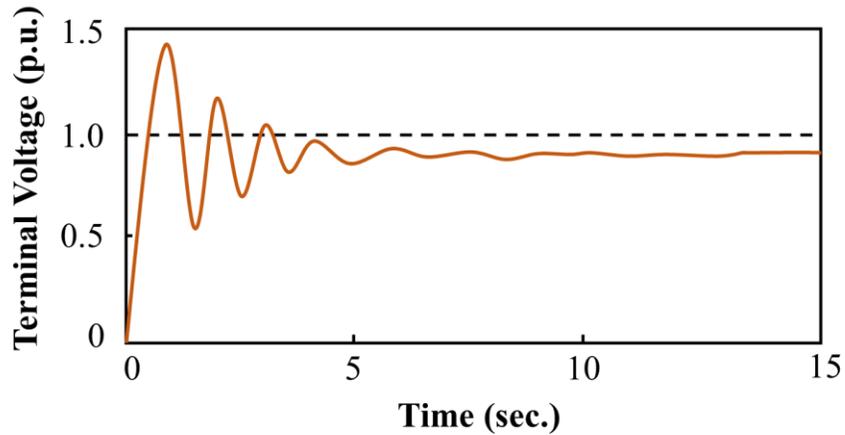

Fig. 6. Terminal voltage variation of an AVR system without any PID controller.

Next, a PID controller is added to the AVR system (AVR-PID). Fig. 7 represents the resulting system block diagram with a step response input. The transfer function of the AVR-PID system is:

$$G_{AVR-PID}(s) = \frac{V_T(s)}{V_{ref}(s)} = \frac{K_A K_E K_G (1+T_S s)(s^2 K_D + s K_P + K_I)}{s(1+T_A s)(1+T_E s)(1+T_G s)(1+T_S s) + K_A K_E K_G K_S (s^2 K_D + s K_P + K_I)} \quad (7)$$

The gain parameters of the PID controller need to be tuned. For this, the classical Ziegler–Nichols (ZN) tuning method is considered. The gain parameters value obtained by the ZN methods at the nominal operating condition ($K_G = 1.0$, $T_G = 1.0$) are: $K_P = 1.0210$, $K_I = 1.8743$ and $K_D = 0.1390$ [43]. Fig. 8 depicts the dynamic terminal voltage response of an AVR-PID system after PID is tuned by ZN methods [42]. The open-loop characteristics parameters and pole-zeros are attached to Table. 2 and Table. 3 for comparison with the AVR system with no controller, respectively [42].

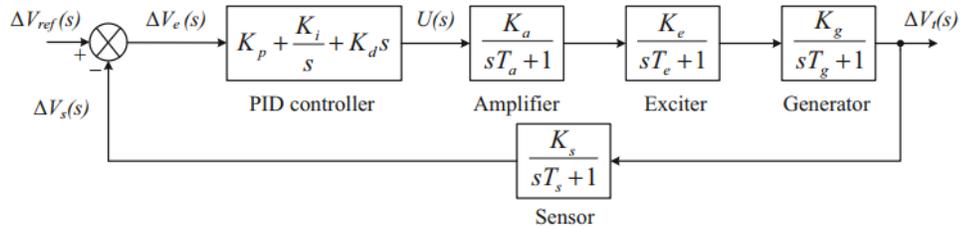

Fig. 7. Block diagram of an AVR-PID transfer system [39].

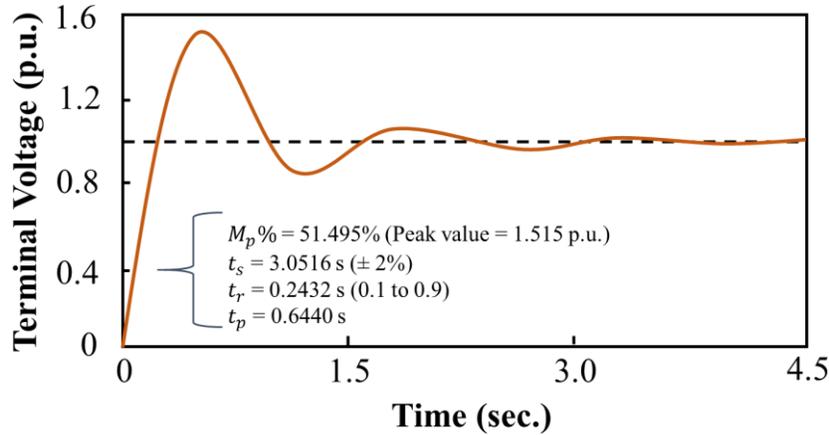

Fig. 8. Terminal voltage variation of an AVR system with ZN-based PID.

It is observed that the ZN-based PID tuning can only slightly improve the system response to a step input voltage. Due to the paucity of the classical ZN method and pole-placement method, modern optimization techniques are considered to improve the dynamic behavior of the system with the help of the PID-controller settings.

Table. 2. AVR system response with and without a controller *[42]*.

| Parameter | Without controller | After using PID controller with ZN tuning |
|---|---|---|
| Mp % | 65.4272% (peak value=1.5037 p.u.) | 51.495 % (peak value = 1.515 p.u) |
| Ts | 6.9711 (± 2%) s | 3.0516 (± 2%) s |
| Tr | 0.2607 s | 0.2366 s |
| Tp | 0.7547 s | 0.6440 s |
| Ess | 0.0907 | 0.0453 s |

Table. 3. AVR system pole-zero locations without a controller *[42]*.

| Zero-poles | Pole-zero locations/values Without controller |
|---|---|
| Real poles | P1 = -98.8170 + 0i<br>P2 = -12.4626 + 0i |
| Complex poles | P3 = -0.5285 + 4.6649i<br>P4 = -0.5285 - 4.6649i |
| Zero | Z1 = -100 + 0i |

### IV. Objective Functions: Parameters Estimation of PID Controller

The performance evaluation of any step response of a system is usually described in terms of time-domain parameters such as percentage peak overshoot ($M_p$%), rise time ($T_r$), settling time ($T_s$), peak time ($T_p$), and steady-state error ($E_{ss}$). In the best system performance, the value of these parameters is minimum for a unit step response. Algorithms consider the task of PID controller gains parameter tuning by dynamically checking the performance at the deemed gain in response to any step input. As a performance criterion for the algorithms, researchers have proposed various objective functions (OFs) that can enumerate the status of $M_p$, $T_r$, $T_s$, $T_p$, and $E_{ss}$ at any PID gain parameters set [47]–[49]. The OFs are defined by numerous weighted parameters on the time domain and frequency domain analysis. Often choosing the weights for time-domain or frequency-domain optimal-tuning PID control becomes very difficult. However, the OFs with time-domain integral error performance criteria are used generally, which consider the error signal between the reference signal and the system output [18]. The commonplace integral performance indices include an Integrated Absolute Error (IAE), Integral Time-Weighted-Absolute-Error (ITAE), Integral of Squared-Error (ISE), and Integral of Time-weighted-Squared-Error (ITSE) [50]. The choice of desired integral performance criterion in the frequency domain bases on the relative advantages and disadvantages of these methods. For instance, IAE and ISE weigh errors uniformly independent of time. Any minimization of the IAE and ISE value provides a relatively small percentage overshoot but a long settling time. The problem of long settling time can be addressed by the ITAE and ITSE. However, these criteria do not comply with the stability boundary, and often their derivation process is complex and time-consuming [51]. The performance criteria are mathematically defined as follows:

$$IAE = \int_0^t |V_{ref} - V_T| dt = \int_0^t |e| dt \qquad (8)$$

$$ITAE = \int_0^t t|V_{ref} - V_T| dt = \int_0^t t|e| dt \qquad (9)$$

$$ISE = \int_0^t |V_{ref} - V_T|^2 dt = \int_0^t |e|^2 dt \qquad (10)$$

$$ITAE = \int_0^t t|V_{ref} - V_T|^2 dt = \int_0^t t|e|^2 dt \qquad (11)$$

where $V_T$ is the terminal voltage. In addition, Zwe-Lee Gaing (ZLG) [14] proposed an objective function that solely considers and weights the transient response parameters ($M_p$, $T_r$, $T_s$, $T_p$, and $E_{ss}$). The ZLG's objection function, W(K), considers only one weight β as:

$$W(K) = Min\{F(K_P, K_I, K_D)\} = (1 - e^\beta)(M_p + E_{ss}) + e^{-\beta}(T_s - T_r) \qquad (12)$$

When β > 0.7, $M_p$ and $E_{ss}$ are minimum. When β < 0.7, $T_r$ and $T_s$ are minimum. Usually, the weight ranges from 0.8 to 1.5. In AVR studies, the majority of the optimization considers the ITSE and ZLG. However, ITSE results in higher $M_p$, and ZLG gives a high value of $T_r$ and $T_s$. This deficit can be subdued by another objective function, J(K) [42], that considers both ITSE and ZLG with a proper weight as:

$$J(K) = Min\{F(K_P', K_I', K_D')\} = (\mu \times ITSE) + ZLG \qquad (13)$$

Here μ is the weight factor. If μ increases, more weight is given to the ITSE; thus, reduced $M_p$ and $E_{ss}$ are obtained. When μ decreases, ZLG got a preference, and therefore, $T_r$ and $T_s$ are minimum. For proper tuning, μ is found to be 30 to 70, with 50 being the best value. Focusing on the application and critical controlling parameters, the objective function can also be formulated as:

$$OF_1 = (W_1 \times ITAE) + (W_2 \times T_S) + (W_3 M_P) \qquad (14)$$
$$OF_2 = (W_1 \times IAE) + (W_2 \times T_S) + (W_3 M_P) \qquad (15)$$
$$OF_3 = (W_1 \times ITSE) + (W_2 \times T_S) + (W_3 M_P) \qquad (16)$$
$$OF_4 = (W_1 \times ISE) + (W_2 \times T_S) + (W_3 M_P) \qquad (17)$$
$$RMSE = \int_0^t \frac{1}{T}\sqrt{|V_{ref} - V_T|^2}\,dt = \int_0^t \frac{1}{T}\sqrt{|e|^2}dt \qquad (18)$$

It is required to perform a quantitative evaluation beforehand to choose the correct OF and weights ($W_1$, $W_2$, and $W_3$) for the PID controller application.

## V. Algorithms for PID Gain Parameters Tuning

Algorithms are used to tune the PID gain parameters and to revert the system to the nominal condition in a fast and efficient manner. In this section, an overview of the proposed heuristic methods for PID tuning in the last decade is accumulated briefly.

### a) Biogeography-Based Optimization (BBO):

Biogeography-Based Optimization (BBO) [52] is an evolutionary algorithm that uses numerical studies focusing on the geographical distribution of the living organisms around a defined area over the course of time [53]. The immigration and emigration of the species to certain habitat depends on the Habitat Suitability Index (HSI) of the habitat and the variables that maintain the habitability, to be defined by the Suitability Index Variables (SIVs) [21]. The PID parameters search procedure with BBO can be processed by [21]:

- Initializing BBO parameters (population size, immigration rate, emigration rate, step size during integration, no of iterations, and others);

- Choose the initial PID gain values satisfying the AVR constraints (SIVs of each habitat);
- Evaluating the fitness value (enumerating the HSI by considering SIV);
- Filter the best-fitted solutions (best value of HSI);
- Applying the probabilistic approach to measure immigration rate to this site by the poorly fitted solutions (non-elite habitat). Calculate new fitted value;
- Update the probability count by considering the solutions. Use mutation for the non-best solutions and calculate the updated HIS;
- Repeat step for the no of iteration size

**b) Gravitational Search Algorithm (GSA):**

Gravitational Search Algorithm (GSA) was first improved by Rashid *et al.,* incorporating the concepts of Newton's laws of gravity and motion [45]. According to [54], in the GSA technique, the agents are coined as objects. The masses of the items are considered via a fitness function to assess their performance. In a net of attractive gravitational force, everything is attracted to the others, and the motion of attraction directs to the agents with heavier mass. Generally, heavy masses are preferred as reasonable solutions. The gravitational constant, best/worst fitness assessment, active-passive and inertia masses, the velocity of the agents, and the position in the search space are mathematically modeled by Duman *et al.* [54]. In Fig. 9, the algorithmic flowchart of the GSA technique is attached.

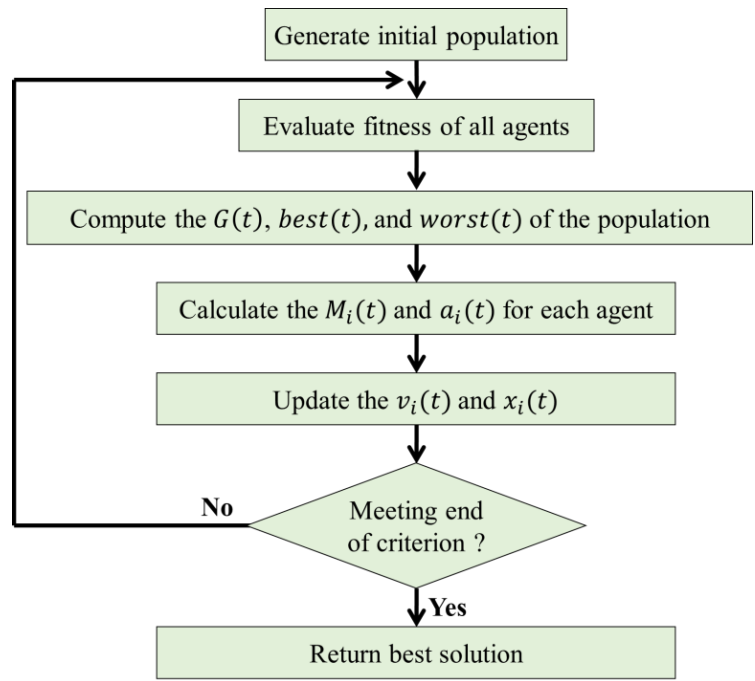

Fig. 9. Flowchart of GSA.

### c) Bacterial Foraging Optimization Algorithm (BFOA):

The concept of bacterial foraging application for distributed control systems was first introduced by Passino *et al.* in 2002 [55]. During the foraging (locating, handling, and ingesting food), the E. coli bacterial performs four steps chemotaxis, swarming, reproduction, and elimination. The bacteria are dispersed onto a search space where each individual bacterium represents a solution. Using effective fitness value, specific operations, and object functions in an iterative way, the optimal solution can be evaluated [32]. The detailed algorithmic steps of the BFOA can be found in [55].

### d) World Cup Optimization (WCO) Algorithm:

In 2016, Razmjooy *et al.* introduced the World Cup Optimization (WCO) algorithm [41]. WCO considers a FIFA tournament and the competitive approach among the countries (teams). According to Fig. 10, the teams within continents compete with each other; the best teams are selected, these teams then compete on the next stage, ultimately resulting in a single best team. A fitness function is used to enumerate the global solution in a continuous optimization procedure by considering at least a global minimum [56]. Mathematical modeling of the continents and their teams' initiation, competing among the teams, and process for the next stage can be found in details in [41].

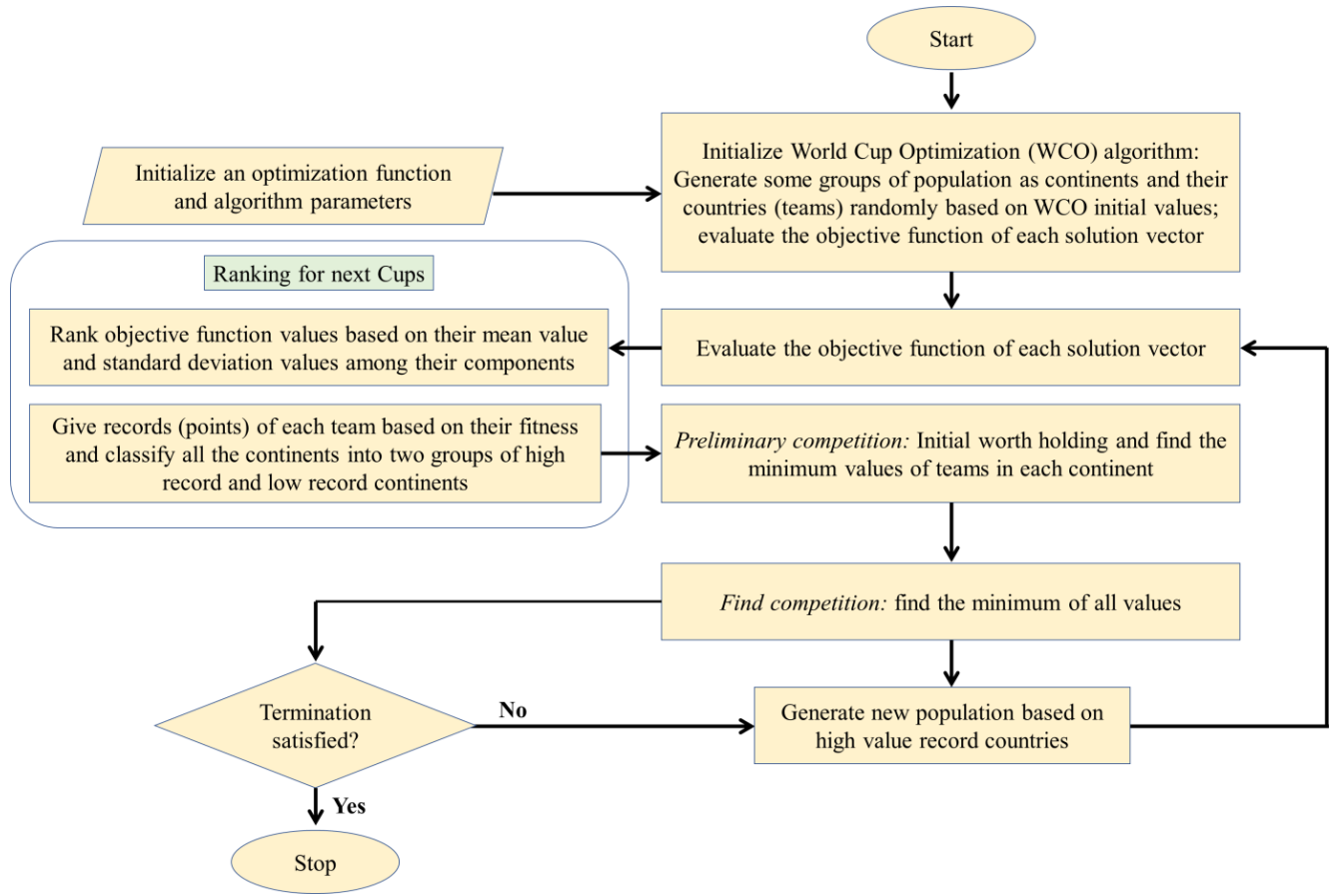

Fig. 10. Algorithmic flowchart of WCO.

e) **Ant Colony Optimization (ACO) Algorithm:**

The Ant Colony Optimization Algorithm (ACO) is a graph-based meta-heuristic algorithm usually employed to solve complex combinatorial optimization problems. ACO considers the food search by an ant for the lowest possible path traveling. In ACO theory, the minimum cost function is obtained by global dissemination and cooperation of path information among the ants [57]. In [17], mathematical modeling of ACO is employed to enumerate the fractional order PID ($PI^\lambda D^\mu$) values. The authors have used ACO and GA to compare the result of the obtained PID gain parameters.

f) **Ant Lion Optimizer (ALO) Algorithm:**

In 2015, Mirjalili introduced the Ant Lion Optimizer (ALO) algorithm, a nature-inspired heuristic algorithm that mimics the catching of ants, one of the antlions' principal prey, by the antlion to solve single and multi-dimensional optimization problems [58]. The hunting nature of antlion was dissected onto the following five major algorithmic steps; the modeling is provided in [31]:

- Random walk of ants around the search space (considers parameters matrices and constraints);
- The roulette wheel method is used to get the fittest antlion that ensures a higher probability of catching prey (optimum solution);
- In each iteration, an ant is caught by the fittest antlion. The random walk of the ants is influenced by the antlions' traps, and random walk are adaptively decreased;
- After the ant is trapped antlion slides to the trapped ant. An ant is captured if the fitness value of the ant is higher than the antlion.
- The antlion changes its position and builds an improved trap for new prey.

g) **Monarch Butterfly Optimization (MBO) Algorithm:**

The monarch Butterfly Optimization (MBO) algorithm was proposed by Wang *et al.* in 2015, inspired by the migration of monarch butterflies from the USA and Southern Canada (land 1) to Mexico (land 2) [59]. MBO uses the migration behavior of the Monarch Butterflies (MBs) for solving global optimization problems. The behavioral steps consist of:

- The total population consists of the MBs located in land one and land 2;
- Migration Operator (MO) engenders the offspring from the total population;
- If the new offspring provides better fitness than their parents, parents are replaced by the newer generation. Else, the offspring is destroyed;
- In each generation, the comparatively better-fitted band is selected;

- The process generates and sets the best items available, and the fitness value remains unchanged by any operator with the increment of the MB population

In the case of a PID controller, the gain parameters are selected by comparing the fitness value of the successive iterations. The pseudo-code and MO's mathematical modeling can be found in [30].

### h) Cuckoo Search (CS) Algorithm:

In 2009, a population-based metaheuristic algorithm to be called the Cuckoo Search (CS) algorithm was first developed by Yang *et al.* 2009 [60]. CS algorithm focuses on the parasitic breeding strategy of certain cuckoos on their way to lay eggs on other birds' nests that the cuckoo perceives as best for raising its' offspring. In the optimization procedure, the main goal is to detect the best nest with optimal solutions, where each individual nest to be considered as a potential solution. Fig. 11 represents the algorithmic flowchart of the CS method. In [18], PID controller gains are optimized by employing detailed mathematical modeling of the CS method.

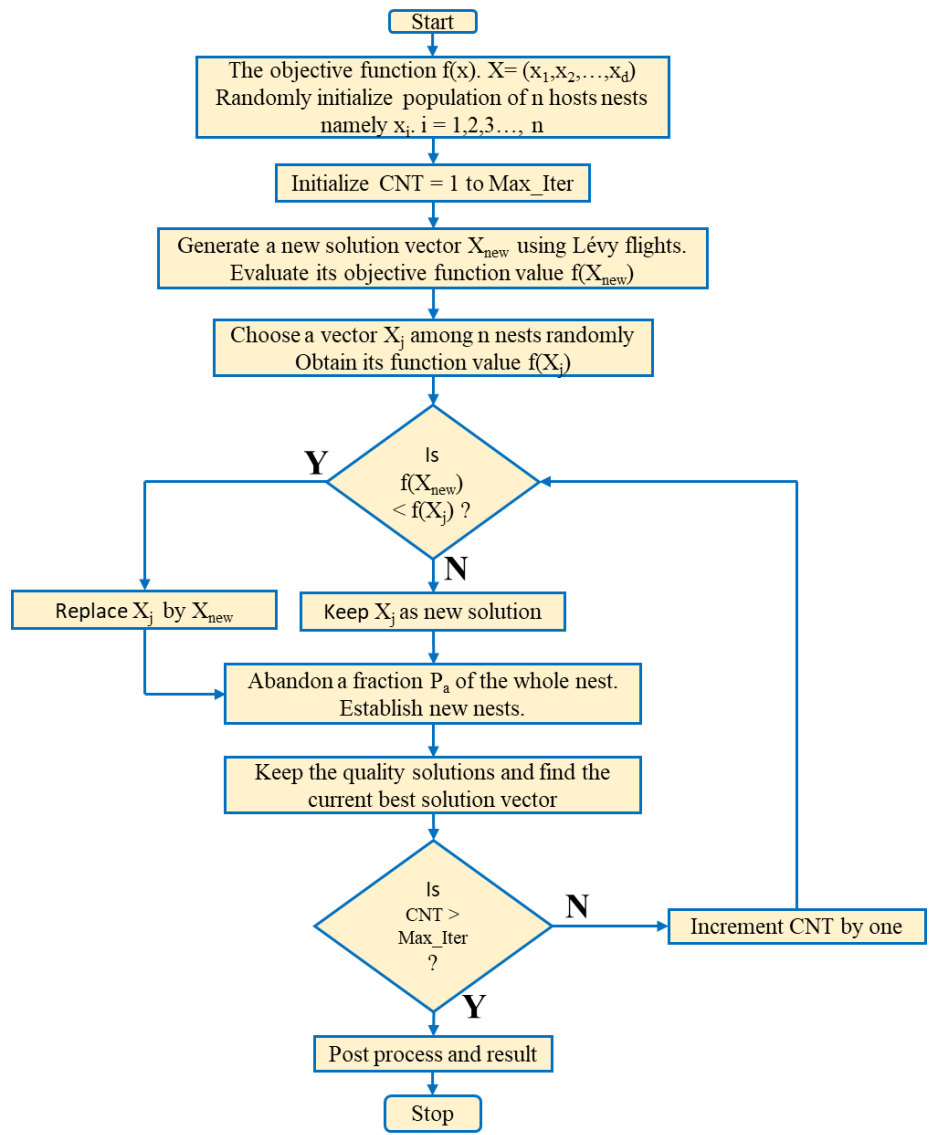

Fig. 11. Algorithmic flowchart of CS algorithm.

### i) Grasshopper Optimization Algorithm (GOA):

Grasshopper Optimization Algorithm (GOA) was proposed by Saremi *et al.* in 2017 [61]. In GOA, the natural behavior of grasshopper swarms is mathematically modeled to solve optimization problems. The attraction and repulsion between grasshoppers inspired to explore search space by repulsion and to exploit optimal region by attraction forces. GOA considers a co-efficient that can modulate the ratio of exploration and exploitation, thus, reduces the local optima trapping and results in a precise global optimum solution. The conceptual model considers the following mathematical relation to updating position during each iteration till the global minima are obtained [26]:

$$X_i^d = r \left( \sum_{\substack{j=1 \\ j \neq i}}^{N} r \frac{(ub_d - lb_d)}{2} s(|x_j^d - x_i^d|) \frac{x_j - x_i}{d_{ij}} \right) + T_d \qquad (30)$$

Where, $X_i^d$ is the position of the current solution in the $d^{th}$ dimension, $r$ refers to a diminishing coefficient, $(ub_d - lb_d)$ is the difference between upper and lower bound in the $d^{th}$ dimension, function $s$ refers to the social forces between two grasshoppers, $(x_j - x_i)$ is the difference between the $j^{th}$ and the $i^{th}$ grasshoppers with an absolute difference $d_{ij}$ between them, and $T_d$ is the best solution value yet achieved for the $d^{th}$ dimension. PID controller parameter tuning using the GOA is described in detail in [26].

### j) Stochastic Fractal Search (SFS) Algorithm:

Stochastic Fractal Search (SFS) algorithm is a natural growth-inspired metaheuristic algorithm first presented by Salimi *et al.* in 2005 [62]. SFS is based on the fractal concept of mathematics and considers three basic steps: initialization, diffusion, and updating. Initially, a few generations are dispersed in the search space for fast convergence and higher accuracy. Then the diffusion step follows, where the fractal search is used for diversifying the present position and fulfilling exploitation. This helps to reduce local minima trapping. Finally, the location of points is updated depending on the positions of other issues in the same group. Çelik *et al.* in [39] constructed the mathematical modeling for finding the optimal gain parameters of a PID controller in an AVR system using the SFS algorithm. The implementation flow diagram is attached in Fig. 12.

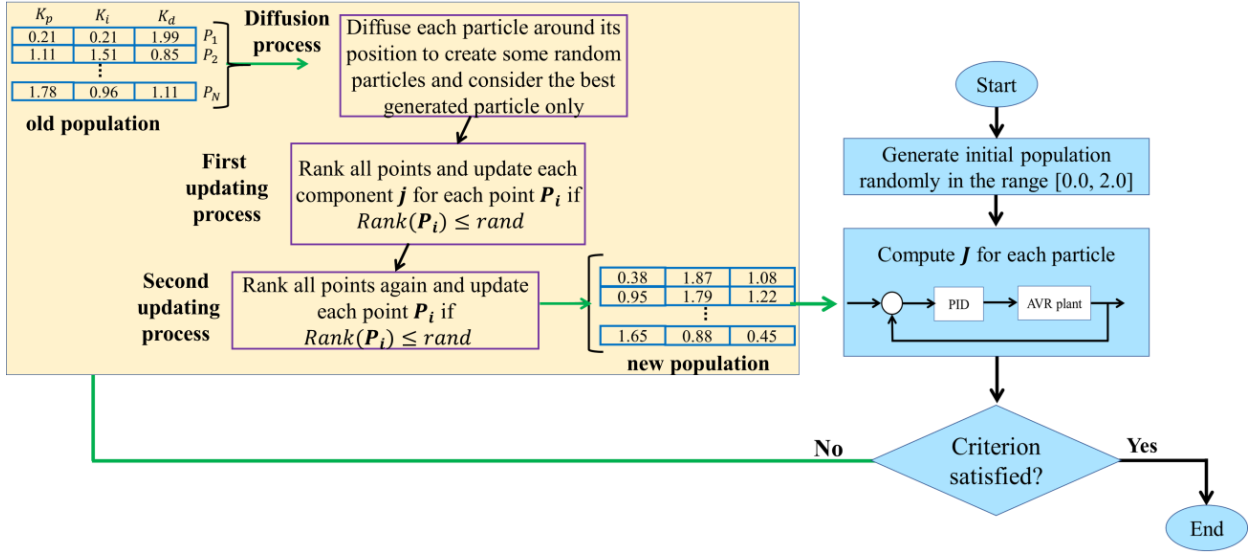

Fig. 12. Implementation of SFS algorithm in an AVR system.

**k) Symbiotic Organisms Search (SOS) Algorithm:**

The symbiotic Organisms Search (SOS) algorithm was first proposed in 2014 by Cheng *et al.* [63]. SOS incorporates the natural procedures of various organisms to survive and grow in the ecosystem. This robust algorithm operates on three primary symbiotic strategies such as mutualism, commensalism, and parasitism. In the symbiotic mutualism interaction, each organism benefits the other; in the case of commensalism, only one of the organism is benefitted while the rest remain neutral, and in parasitism, one of the organism is benefitted by placing parasitic effect on the other, causing the later one to be affected [64]. An organism in the search space utilizes all these three kinds of interactions randomly with other organisms in the ecosystem. After completing all these strategies, SOS tries to generate another generation of the organism that will result in a feasible cost function. The procedure then repeats till the optima are obtained. The SOS is employed to extract the optimum points for the PID controller gains in an AVR system modeled by Çelik *et al.* [33]. Fig. 13 represents the procedure for the SOS implementation for optimizing PID gain parameters in an AVR system.

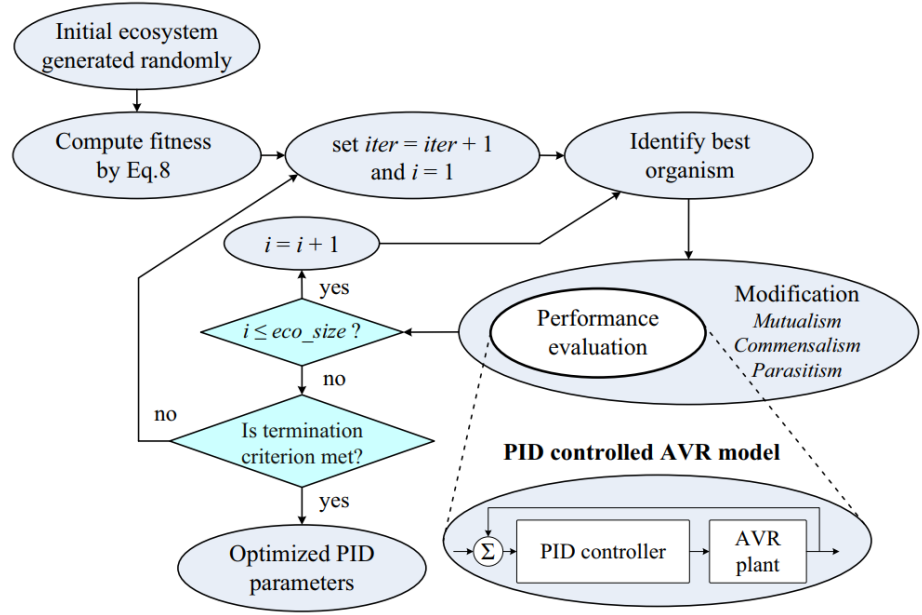

Fig. 13. Implementation of SOS algorithm in an AVR system *[33]*.

l) **Salp Swarm Algorithm (SSA):**

In 2018, Ekinci *et al.* proposed the Salp Swarm Algorithm (SSA) inspired by salps' natural propagation and behavior [29]. The transparent bodied salps are renowned for their spiral chain formation during movement and food searching. In SSA, a leader and followers need to be defined to lead the chain to the food source (global minima). In this progression, the salp explores and exploits the entire search space. A range of salps (agents) is primarily spread in all the search space. The assessment of the current population of salps then defines the best salp and makes it the leader. Other salps follow this salp until another leader emerges in the progressive iterations. In [65], the SSA is incorporated to model and optimize Fractional Order PID (FOPID) controller parameters. The algorithmic flowchart of the procedure is shown in Fig. 14.

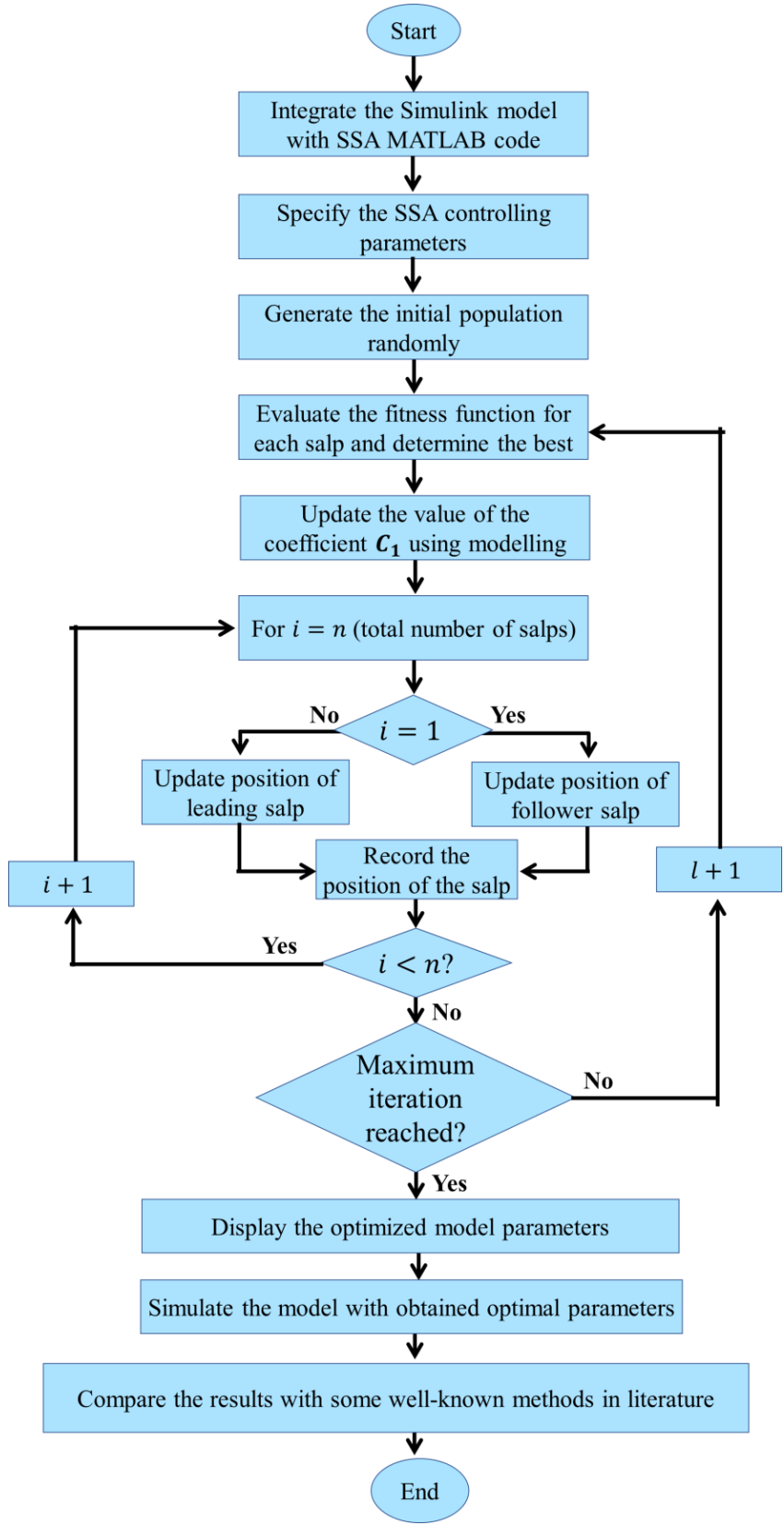

Fig. 14. SAS-based FOPID controller parameter selection in an AVR system.

**m) Sine Cosine Algorithm (SCA):**

Sine Cosine Algorithm (SCA) is a population-based metaheuristic optimization algorithm first proposed by Mirjalili in 2016 [66]. SCA initializes various random solutions in the search spaces and then forces them to move onwards or outwards the best solution. Moreover, several adaptive and random variables are often integrated into the algorithm to emphasize the exploitation and exploration of search space. In [67], Hekimoğlu mathematically modeled an AVR system by optimizing the SCA's PID gain parameters. Fig. 15 represents the SCA implementation in an AVR system.

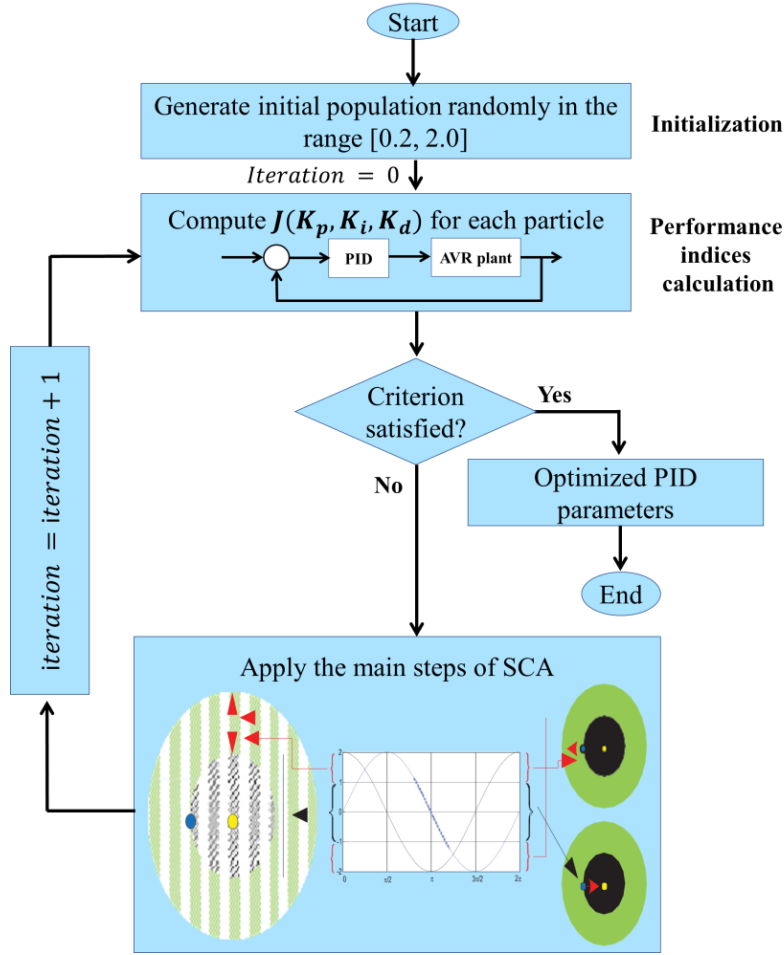

Fig. 15. SCA-based PID gain optimization in an AVR system.

**n) Improved Kidney-inspired Algorithm (IKA):**

In 2017, Jaddi *et al.* first proposed the Kidney-inspired Algorithm (KA) mimicking the filtration, reabsorption, secretion, and excretion processes of the kidney to solve dynamic optimization problems [68]. In the KA technique, an arbitrary populace (generation) of possible solutes (solution) is considered. The solutes in each iteration process through the fore-mentioned four stages. At the end of the process, a new solute is generated by moving towards the observed best solutes [69]. Successive iteration ultimately results in the best global optimal solution to comply with the fitness value and objective functions. Detailed mathematical modeling of IKA-based PID controller parameters can be found in [42]. In Fig. 16, the technique IKA implementation for the AVR system is summarized.

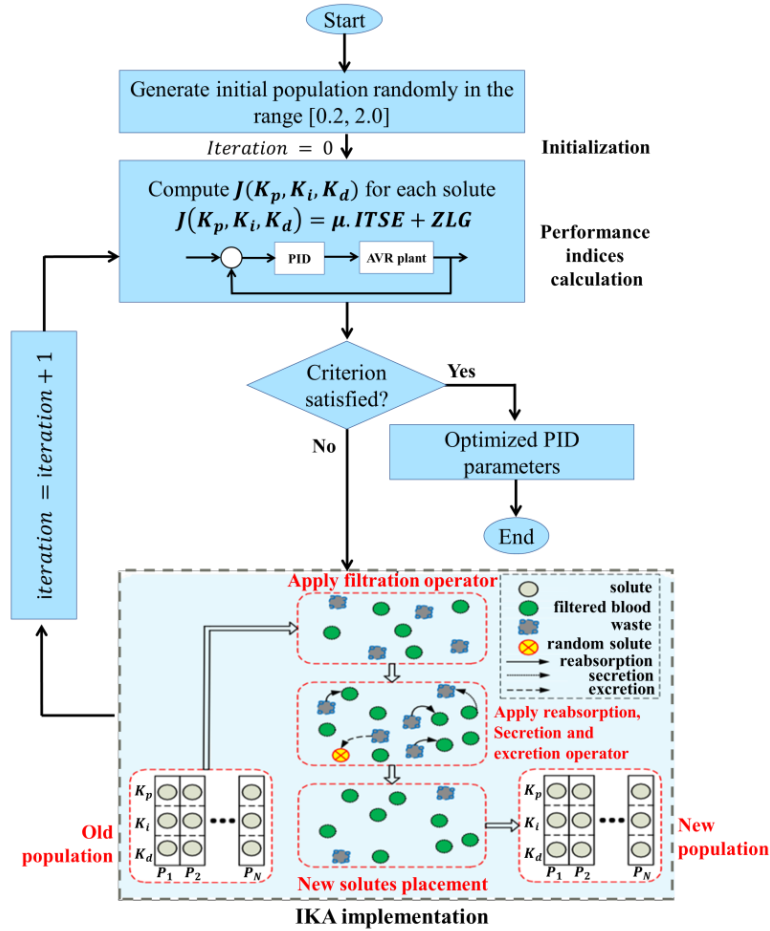

Fig. 16. Implementation of IKA algorithm in an AVR system.

o) **Tree-Seed algorithm (TSA):**

Tree-Seed Algorithm (TSA) is a nature-inspired population-based metaheuristic algorithm first proposed by Kiran in 2015 [46]. In TSA, the tree and the seeds act as solutions, with the area under consideration being the search space. The trees are randomly dispersed in the exploration phase. Then, the seeds demonstrating closer characteristics of the trees are considered in the exploitation phase. A greedy selection function is used to find out the best seeds of each tree. This recursive process continues till it is converged. In [43], detailed mathematical modeling of an AVR optimization with a TSA algorithm is provided. The critical optimization processes are incorporated, as shown in Fig. 17.

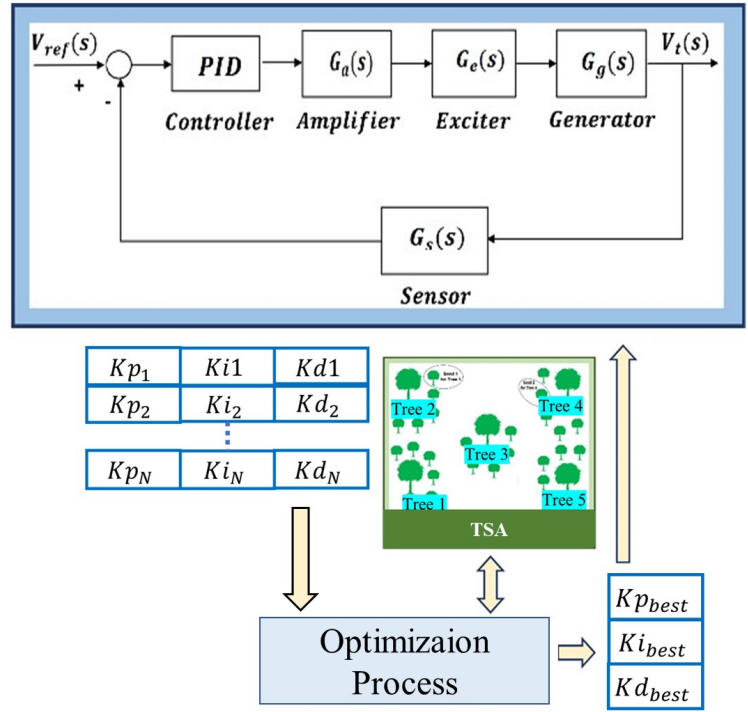

Fig. 17. Implementation of TSA algorithm in an AVR system *[43]*.

## VI. Algorithms for PID Gain Parameters Tuning

A comparative analysis among the majority of the previous optimization algorithms' output in an AVR system is demonstrated. It is seen that all the previous investigation follows a general trend while evaluating the performance of their considered algorithm. Before presenting any significant result, researchers outlined the simulation platform and the computing device's specifications that were being used throughout the simulation. Apart from a few, all the investigations used MATLAB/Simulink toolbox to model the AVR system. The iteration size, population size, and other required variables (algorithm-specific) are defined for establishing a simulation preset.

During the simulation, various algorithms have incorporated a diverse amount of objective function and performance measuring criterion. The simulation is conducted till the OF results in the minimum convergence characteristics. It is essential that due to penalty co-efficient and different OF, the algorithms will result differently. Moreover, comparative performances among algorithms are only feasible if the algorithms are simulated under the same computing device, population size, and objective functions.

For lucid perception, in Table. 4 majority of the prominent AVR optimization algorithms with their objective function, transient response parameters, pole-zero mapping, and bode diagram analysis are accumulated. Though there is some slight variations in the model defined in various research in terms of the gain value of amplifier, exciter, rectifier, generator, and sensor, a pattern can be drawn from the observation of Table. 4. The majority of the previous algorithm have used ITSE, ITAE, and their variations as their objective function. Other OF includes RMSE, modified ITAE, FOD, and user-defined aim of custom functions. It is clear that RMSE results in higher PID parameter values, followed by ITSE, ITAE, and Figure of Demerit (FOD). The range of gain parameters for the ITAE and ITSE objective functions are addressed explicitly in Table. 5. With ITSE OF, ALO renders the lowest gains for proportional and integral gains. ACO and PSA result in the lowest and highest gain values with ITAE OF for all three gains, respectively.

For the optimized PID gains, it is required that such a set of values results in good transient performance. A unit step input was provided to the optimized model, and the terminal voltage results are noted for the methods. The AVR system's terminal voltage profile with the eight most popular metaheuristic algorithms is given in Fig. 18. It can be observed that the recently proposed TSA gives a better voltage profile than the others [43]. Table. 6 outlines the time-domain performance with ITSE and ITAE objective functions. The SFS algorithm provides the lowest rise time and settling time with ITSE OF. The ALO results in nil maximum peak overshoot with the burden of the highest settling time. However, with ITAE OF, CFA gives the nil maximum overshoot and the lowest settling time. The most down rise time comes with PSA.

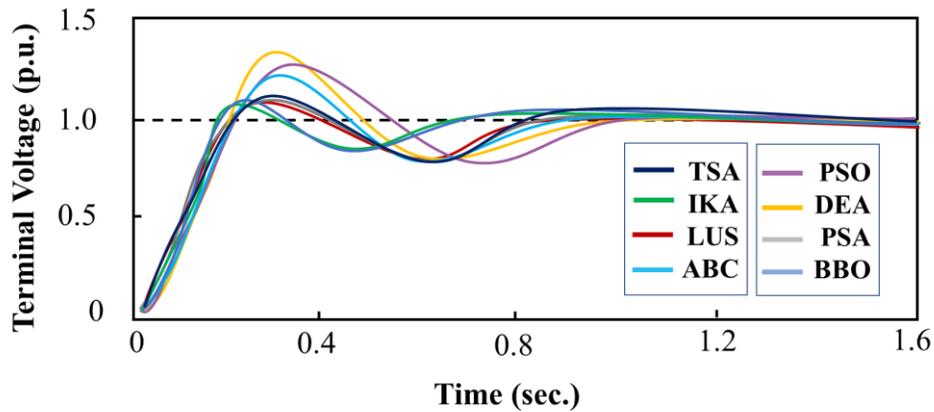

Fig. 18. AVR terminal voltage profile with eight popular algorithms *[43]*.

After considering the optimized AVR system, the pole-zero mapping is required to calculate the damping ratio and system response by dint of the root-locus plot. Table. 5 also demonstrates the closed-loop poles with a relative damping ratio. All the poles (real and complex) are in the left half of the s-plane, thus, ensures the stability of the resultant system.

The bode analysis is also popular while evaluating the performance of the control system algorithms. According to Table. 5, the highest peak margin was shown by the DE algorithm to be 4.2 dB. For SOS, LUS, MOL, and ALO, about 180° phase margin and infinite delay margin are achievable. KA and SOS provide the maximum and the minimum bandwidth (BW) values of 15.11 Hz and 6.15 Hz.

The dynamic response is also graded by the dint of performance's robustness during any change of the amplifier, sensor, generator, or exciter gain parameters. In various investigations, robustness is evaluated by changing the gains ($T_A$, $T_E$, $T_G$, and $T_s$) from -50% to 50% with a step size of 25% and then analyzing the transient performance characteristics ($T_r$, $T_s$, $M_p$, $E_{ss}$, and $T_p$). During any uncertain parameters in the system, this method acts as a rule of thumb to investigate discrepancy performance efficiently.

With any variation in the connected load of the synchronous generator, the gain parameters change. The robustness and transient responses of the controllers during any dynamic load fluctuation thus need to be observed. The disturbance responses of an AVR system are depicted in Fig. 19. Here, two disturbances are added at 10% during the 3$^{rd}$ second and 5$^{th}$ second. The algorithms revert the system to its nominal terminal voltage in a fast manner. In terms of robustness, both the IKA and TSA dominates all the other major metaheuristic algorithms.

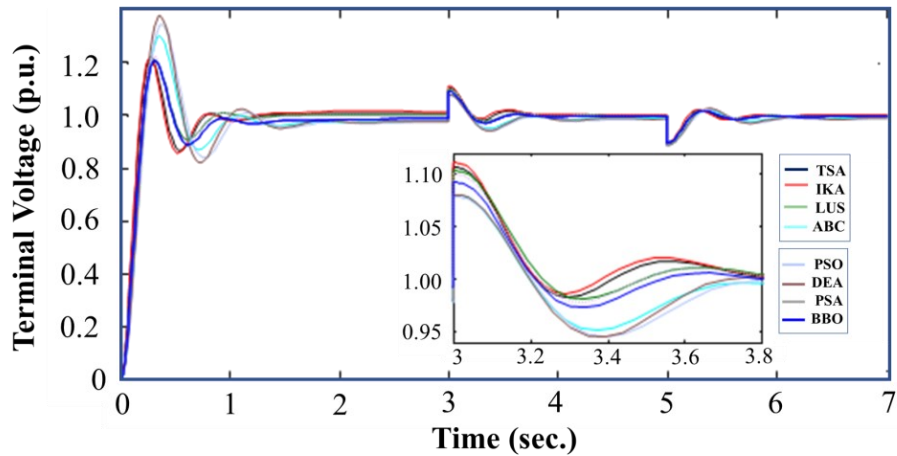

Fig. 19. AVR disturbance response for eight popular algorithms *[43]*.

## VII. AVR optimization with metaheuristic algorithms

During the last decade, metaheuristic algorithms faced significant growth in their robustness, fast computation, and early convergence. From Table. 7, it can be perceived that from 2011, there has been a dominant improvement in the field of PID controlling by metaheuristic algorithms. Various tuning methods, performance metrics, and analysis methods were performed to compare each new algorithm with one or more than one previously proposed algorithm. Due to the shortcomings of classical pole-zero placement or ZN methods to improve the PID behavior, more unique heuristic optimization techniques were proposed. Objective functions including ITSE, ITAE, IAE, ISE, ZLG, and custom OF have been used to simulate any regular and disturbing load dynamics. Transient responses, pole-zero mapping, robustness analysis, and disturbance analysis enumerated the proposed algorithms' significance. In Table. 8, algorithms addressed in this paper are tabulated by accumulating utilized objective function(s), time-domain performance matrices, frequency domain responses, and observed PID gain values.

Though researchers worldwide have tremendous participation worldwide to gain an optimization algorithm for AVR system that will cost the minimum computation burden and time while providing satisfactory performance, the actual progress on the field is yet vague. The majority of the algorithms proposed depends on various parameters starting from the device used, its memory size, number of considered iteration and population size, and many more. While presenting a new algorithm, it is expected to compare all the previously proposed algorithms in the same milieu to outline the new algorithm's actual contribution. However, due to computational burden and lack of time, such has seldom been attached to any papers. Moreover, the current trend to use custom objective function needs to be analyzed closely with previous investigations so that any actual breakthrough can easily be understood.

Table. 4. Optimized PID gains with time domain and frequency domain response of the AVR system.

| Proposed Algorithm | Objective Function | PID parameters (optimized) | | | Time-domain analysis | | | | | | Bode Analysis | | | | | |
|---|---|---|---|---|---|---|---|---|---|---|---|---|---|---|---|---|
| | | $K_P$ | $K_I$ | $K_D$ | Maximum Peak Overshoot (p.u) | Peak Time (sec) | Overshoot (%) | Rise Time (sec) | Settling Time (sec) | $E_{ss}$ | Close-loop Poles | Damping Ratio | Peak Gain (dB) | Phase Margin (deg.) | Delay Margin (s) | BW (Hz) |
| GA [56] | RMSE | 6.71275 | 3.2230 | 14.36576 | - | - | 49.9 | 0.755 | 6.92 | - | - | - | - | - | - | - |
| PSO [14] | ITSE | 1.7774 | 0.3827 | 0.3184 | 1.299 | 0.374 | 29.92 | 0.163 | 3.399 | - | -100.85 -3.09 ± 7.80i -6.26 -0.22 | 1.00 0.36 1.00 1.00 | 3.75 | 62.2 | 0.103 | 12.182 |
| COLM [38] | Modified-ITAE | 0.8870 | 0.6460 | 0.3090 | - | - | - | - | - | - | - | - | - | - | - | - |
| CAS [27] | ITAE | 0.5613 | 0.3670 | 0.2326 | 0.4653 | - | - | 0.3028 | 0.4434 | 0 | - | - | - | - | - | - |
| ABC [15] | ITSE | 1.6524 | 0.4083 | 0.3654 | 1.248 | 0.375 | 24.85 | 0.157 | 3.093 | - | -100.98 -3.76 ± 8.41i -4.75 -0.25 | 1.00 0.40 1.00 1.00 | 2.87 | 69.4 | 0.1109 | 12.8791 |
| MOL [19] | ITSE | 0.9877 | 0.7780 | 0.5014 | 1.1143 | 0.3056 | - | 0.1468 | 0.8769 | - | -100.0 -4.92 ± 4.72i -2.11 -1.06 | 1.00 0.72 1.00 1.00 | 0.0 | 180 | Inf. | 6.3373 |
| ASO [35] | ITAE | 0.9068 | 0.4626 | 0.3279 | - | - | - | - | - | - | - | - | - | - | - | - |
| PSA [36] | ITAE | 1.2771 | 0.8471 | 0.4775 | 1.169 | 0.316 | 16.84 | 0.144 | 0.804 | - | -101.0 -1.28 ± 0.15i -4.82 ± 10.1i | 1.00 0.99 0.43 | 2.87 | 69.4120 | 0.1109 | 12.8791 |
| CFA [28] | ITAE | 0.6610 | 0.4951 | 0.2561 | 0 | - | - | 0.2691 | 0.3739 | 0.0051 | - | - | - | - | - | - |
| LUS [25] | Modified-ITSE | 1.2012 | 0.9096 | 0.4593 | 1.154 | 0.313 | 15.46 | 0.151 | 0.799 | - | -101.25 -1.24 ± 0.56i -4.88 ± 9.88i | 1.00 0.91 0.44 | 0.0 | 180 | Inf. | 7.1667 |
| TLBO [2] | FOD | 0.5302 | 0.4001 | 0.1787 | - | - | - | 0.3537 | 0.5603 | - | - | - | - | - | - | - |
| BBO [21] | ITSE | 1.2464 | 0.5893 | 0.4562 | 1.155 | 0.317 | 15.52 | 0.149 | 1.446 | - | -100.0 -4.80 ± 10.2i -2.1 -0.585 | 1.00 0.43 1.00 1.00 | 1.56 | 81.6 | 0.122 | 14.284 |
| GSA [54] | ITSE | 1.4379 | 1.2208 | 0.7363 | 1.240 | 0.24 | - | 0.107 | 0.597 | - | - | - | - | - | - | - |
| BFOA [32] | Custom OF | 1.087 | 0.83064 | 0.4077 | 1.174 | 0.335 | - | 0.154 | 0.759 | - | -101.12 -4.94 ± 9.18i | 1.00 0.47 0.91 | 0.965 | 91.45 | 0.156 | 13.04 |

| | | | | | | | | | | | | | | | |
|---|---|---|---|---|---|---|---|---|---|---|---|---|---|---|---|
| | | | | | | | | | | -1.26 ± 0.56i | | | | | |
| WCO [41] | ITAE | 0.8130 | 0.1787 | 0.2205 | - | - | - | - | 2.94 | - | - | - | - | - | - | - |
| ACO [17] | ITAE | 0.2442 | 0.1423 | 0.0427 | - | - | 8.871 | 0.838 | 2.641 | 0 | - | - | - | - | - | - |
| ALO [31] | ITSE | 0.81918 | 0.24366 | 0.96832 | 0 | - | - | 0.1384 | 4.236 | - | -101.2 -9.62 ± 4.22i -1.33 -0.804 | 1.00 0.92 1.00 1.00 | 0.0 | 180 | Inf. | 6.58 |
| MBO [30] | ISE | 1.2454 | 0.7516 | 0.4821 | - | - | - | 1.116 | 4.995 | - | - | - | - | - | - | - |
| CS [18] | Custom OF | 0.6198 | 0.4165 | 0.2127 | 0 | - | - | 0.308 | 0.426 | 0 | - | - | 0 | 180 | Inf. | 7.342 |
| GOA [26] | ITSE | 1.3825 | 1.4608 | 0.5462 | 1.205 | 0.286 | - | 0.130 | 0.971 | - | - | - | - | - | - | - |
| SFS [39] | ITSE | 1.2837 | 1.3392 | 0.7780 | 1.228 | 0.230 | - | 0.103 | 0.584 | - | - | - | - | - | - | - |
| SOS [33] | Modified-ITAE | 0.5693 | 0.4097 | 0.1750 | 1.013 | - | - | 0.353 | 0.485 | 0.002 | -100.48 -4.97 ± 4.69i -1.98 -1.10 | 1.00 0.73 1.00 1.00 | 0.0 | 180 | Inf. | 6.15 |
| SSA [29] | ITSE | 1.3381 | 1.1204 | 0.6361 | - | 0.263 | 20.30 | 0.119 | 0.690 | - | - | - | - | - | - | - |
| SCA [67] | Time domain | 0.9826 | 0.8337 | 0.4982 | 1.114 | 0.304 | - | 0.148 | 0.724 | - | -101.37 -5.16 ± 10.52i -0.91 ± 0.82i | 1.00 0.44 0.74 | 1.09 | 87.3 | 0.128 | 14.821 |
| IKA [42] | ITSE | 1.0426 | 1.0093 | 0.5999 | 1.150 | 0.269 | 15.00 | 0.128 | 0.753 | - | -102.0 -5.13 ± 11.7i -0.80 ± 0.93i | 1.00 0.40 0.65 | | | | |
| TSA [43] | ITSE | 1.1281 | 0.9567 | 0.5671 | 1.155 | 0.278 | 15.57 | 0.131 | 0.758 | - | -101.55 -5.05 ± 11.32i -0.93 ± 0.82i | 1.00 0.74 0.40 | | | | |
| DE [37] | ITSE | 1.9499 | 0.4430 | 0.3427 | 1.327 | 0.359 | 32.754 | 0.154 | 2.650 | - | -100.91 -3.03 ± 8.19i -6.30 -0.23 | 1.00 0.34 1.00 1.00 | 4.20 | 58.4 | 0.092 | 12.8 |
| TCGA [34] | Custom OF | - | - | - | - | - | - | 0.63 | 0.52 | 0.00003 | - | - | - | - | - | - |
| KA [69] | ITSE | 1.0685 | 1.0018 | 0.5103 | 1.136 | 0.308 | - | 0.143 | 0.771 | - | -101.4 -5.09 ± 10.6i -0.96+0.92i | 1.0 0.43 0.72 | 1.33 | 83.5 | 0.118 | 15.11 |

Table. 5. PID gain values with ITSE/ITAE objective functions for different algorithms.

| Objective Function | Algorithms used the OF | PID gain | Range | Algorithm for the lowest value | Algorithm for the highest value |
|---|---|---|---|---|---|
| ITSE | ALO [31], SFS [39], GSA [54], SSA [29], IKA [42], TSA [43], GOA [26], KA [69], MOL [19], LUS [25], BBO [21], ABC [15] DE [37], PSO [14] | $K_P$ | 0.81918 - 1.9499 | ALO [31] | DE [37] |
| | | $K_I$ | 0.24366 - 1.4608 | ALO [31] | GOA [26] |
| | | $K_D$ | 0.31840 - 0.9683 | PSO [14] | ALO [31] |
| ITAE | PSA [36], ASO [35], COLM [38], CFA [28], CAS [27], WCO [41], SOS [33], ACO [17] | $K_P$ | 0.24420 - 1.2771 | ACO [17] | PSA [36] |
| | | $K_I$ | 0.14230 - 0.8471 | ACO [17] | PSA [36] |
| | | $K_D$ | 0.04270 - 0.4775 | ACO [17] | PSA [36] |

Table. 6. Transient performances with ITSE/ITAE objective functions for different algorithms.

| Objective Function | Algorithms used the OF | Transient parameter | Range | Algorithm for the lowest value | Algorithm for the highest value |
|---|---|---|---|---|---|
| ITSE | ALO [31], SFS [39], GSA [54], SSA [29], IKA [42], TSA [43], GOA [26], KA [69], MOL [19], LUS [25], BBO [21], ABC [15] DE [37], PSO [14] | $M_p$ (p.u) | 0 - 1.327 | ALO [31] | DE [37] |
| | | $T_r$ (s) | 0.103 - 0.163 | SFS [39] | PSO [14] |
| | | $T_s$ (s) | 0.584 - 4.236 | SFS [39] | ALO [31] |
| ITAE | PSA [36], ASO [35], COLM [38], CFA [28], CAS [27], WCO [41], SOS [33], ACO [17] | $M_p$ (p.u) | 0 - 0.4653 | CFA [28] | CAS [27] |
| | | $T_r$ (s) | 0.144 - 2.94 | PSA [36] | ACO [17] |
| | | $T_s$ (s) | 0.374 - 2.94 | CFA [28] | WCO [41] |

Table. 7. Prominent metaheuristic algorithms for AVR optimization (2001-2020).

| Year of publication | Published algorithms for PID gain tuning in an AVR system | Total count |
|---|---|---|
| 2001 | GA [56] | 1 |
| 2004 | PSO [14], | 1 |
| 2009 | COLM [38], CAS [27] | 2 |
| 2011 | ASO [35], ABC [15] | 2 |
| 2012 | MOL [19], PSA [36] | 2 |
| 2013 | TCGA [34] | 1 |
| 2014 | CFA [28], LUS [25], TLBO [2] | 3 |
| 2015 | PBO [70] | 1 |
| 2016 | BBO [21], GSA [54], BFOA [32], WCO [41], ACO [17] | 5 |

| 2017 | ALO [31], MBO [30] | 2 |
| 2018 | CS [18], GOA [26], SFS [39], SOS [33], SSA [29] | 5 |
| 2019 | SCA [67], KA [69], IKA [42], HHO [71] | 4 |
| 2020 | TSA [43], ASO [72] | 2 |

Table. 8. Tuning algorithms used for AVR systems in the current literature [42]

| | | | BBO [21] | GSA [54] | GOA [26] | PBO [70] | CFA [28] | LUS [25] | TLBO [2] | TCGA | MOL [19] | ASO [35] | PSA [36] | ABC [15] | COLM | CAS [27] | PSO [14] | SFS [39] S | SOS [33] | BFOA [32] | ALO [31] | SCA [67] | SSA [29] | WCO [41] | MBO [30] | CS [18] | IKA [42] | TSA [43] |
|---|---|---|---|---|---|---|---|---|---|---|---|---|---|---|---|---|---|---|---|---|---|---|---|---|---|---|---|---|
| | | Proposed Algorithm in reference | | | | | | | | | | | | | | | | | | | | | | | | | | | |
| Tuning Methods | Classical | ZN | | | ▽ | | | | | | | | | | | | | | | | | | ▽ | ▽ | | | | ▽ | |
| | Heuristic Optimization | ABC | ▽ | ▽ | ▽ | | | ▽ | | | ▽ | | ▽ | | | | | | ▽ | ▽ | ▽ | ▽ | ▽ | ▽ | | ▽ | | ▽ | ▽ |
| | | ALO | | | | | | | | | | | | | | | | | | | | ▼ | | | | | | | |
| | | ASO | | | | | | | | | | ▼ | | | | | | | | | | | | | | | | | |
| | | BBO | ▼ | | | | | | | | | | | | | | | | ▽ | ▽ | | ▼ | ▽ | | | | | ▽ | ▽ |
| | | BFOA | | | | | | | | | | | | | | | | | | | | ▼ | ▽ | | | | ▽ | | |
| | | CAS | | | | | | | | | | | | | | | ▼ | | | | | ▽ | | | ▽ | | | | |
| | | COLM | | | | | | | | | | | | | | ▼ | | | | | | | | | | | | | |
| | | CPSO | | | | | | | | | | | | | | | | | | | | | | | ▽ | | | | |
| | | CS | | | | | | | | | | | | | | | | | | | | | | | | | ▼ | | |
| | | DE | ▽ | ▽ | ▽ | | | ▽ | | | ▽ | | | ▽ | | | | | | | ▽ | ▽ | ▽ | | | | | ▽ | ▽ |
| | | CFA | | | | | ▼ | | | | | | | | | | | | | | | | | | | | | | |
| | | GA | | | | ▽ | | | ▽ | | | | | | | ▽ | ▽ | | | | | ▽ | | | | | ▽ | | |
| | | GOA | | | ▼ | | | | | | | | | | | | | | | | | | | | | | | | ▽ |
| | | GSA | | ▼ | | ▼ | | | | | | | | | | | | | ▽ | | | | | | | | | | |
| | | GWO | | | | | | | | | | | | | | | | | | ▽ | | | | | | | | | |
| | | ICA | | | | | | | | | | | | | | | | | | | | | | ▽ | | | | | |
| | | KA | | | | | | | | | | | | | | | | | | | | | | | | | | | |
| | | LUS | | | | | | ▼ | | | | | | | | | | | ▽ | ▽ | | | | | | | ▽ | ▽ | ▽ |
| | | MBO | | | | | | | | | | | | | | | | | | | | | | | | ▼ | | | |
| | | MOL | | | | | | | | | ▼ | | | | | | | | ▽ | | ▽ | ▽ | | | | ▽ | ▽ | | |
| | | PSA | | | | | | | | | | | ▼ | | | | | | | | | | | | | | | ▽ | ▽ |
| | | PSO | ▽ | ▽ | | ▽ | ▽ | ▽ | | ▽ | ▽ | ▽ | | ▽ | | | | ▼ | | ▽ | ▽ | | | ▽ | ▽ | ▽ | ▽ | ▽ | ▽ |
| | | SCA | | | | | | | | | | | | | | | | | | | | | | | | | | | |
| | | SFS | | | | | | | | | | | | | | | | | ▼ | | | | | | | | | | |
| | | SOS | | | | | | | | | | | | | | | | | | ▼ | | | | | | | | | |
| | | SSA | | | | | | | | | | | | | | | | | | | | | | ▼ | | | | | |
| | | TCGA | | | | | | | | | ▼ | | | | | | | | | | | | | | | | ▽ | | |
| | | TLBO | | | | | | | | ▼ | | | | | | | | | | ▽ | | | | | | | | | |
| | | WCO | | | | | | | | | | | | | | | | | | | | | | | ▼ | | | | |
| | | IKA | | | | | | | | | | | | | | | | | | | | | | | | | | ▼ | ▽ |
| | | TSA | | | | | | | | | | | | | | | | | | | | | | | | | | | ▼ |
| | | IAE | | ▼ | | | ▼ | | | | ▼ | | | | | | | | | | | ▼ | | | ▼ | | | | ▼ |

| Category | Method | | | | | | | | | | | | | | | | | | | | | | | | | | |
|---|---|---|---|---|---|---|---|---|---|---|---|---|---|---|---|---|---|---|---|---|---|---|---|---|---|---|---|
| Performance Metrics | ZLG | | ▼ | | | ▼ | | | | | | | | | ▼ | | | ▼ | | ▼ | | | | | ▼ | | |
| | ITAE | | ▼ | | | ▼ | ▼ | ▼ | | | ▼ | ▼ | | | ▼ | | | | | ▼ | | | ▼ | ▼ | | | ▼ |
| | ISE | | ▼ | | | | ▼ | | | | ▼ | | | | | | | | | ▼ | | | | ▼ | | | ▼ |
| | ITSE | ▼ | ▼ | ▼ | | | ▼ | | | ▼ | | | | | ▼ | | | ▼ | | | ▼ | | | | | ▼ | ▼ |
| Investigation Methods | Other | | | | ▼ | | | | ▼ | | ▼ | | | ▼ | ▼ | | | ▼ | | ▼ | | | | | | | |
| | Pole-zero map | ▼ | | | | | ▼ | | | ▼ | | ▼ | ▼ | | | | ▼ | ▼ | ▼ | ▼ | ▼ | ▼ | | | | | ▼ | ▼ |
| | Bode plot | ▼ | | | | | ▼ | | | ▼ | | ▼ | ▼ | | | | ▼ | | ▼ | ▼ | ▼ | ▼ | | | | | ▼ | ▼ |
| | Robustness | ▼ | ▼ | ▼ | | | ▼ | ▼ | | | ▼ | ▼ | ▼ | ▼ | | ▼ | ▼ | | ▼ | ▼ | ▼ | ▼ | ▼ | | ▼ | | ▼ | ▼ | ▼ |
| | Transient | ▼ | ▼ | ▼ | ▼ | ▼ | ▼ | ▼ | ▼ | ▼ | ▼ | ▼ | ▼ | ▼ | ▼ | ▼ | ▼ | ▼ | ▼ | ▼ | ▼ | ▼ | ▼ | ▼ | ▼ | ▼ | ▼ | ▼ |

▼ The proposed model  ∇ Method used for comparison

The following aspects should be considered during the future investigation of the metaheuristic algorithm for AVR systems' optimization:

- The algorithmic steps should be reproducible to be quickly established in future investigations with any newer algorithm.
- Detailed mathematical modeling of the objective function and the recursion selection should be well addressed.
- In addition to using any custom objective function, a general objective function should be followed to bring all the algorithms' performance on the same page for comparison.
- The computational device should be mentioned with metrics for change in system optimization and simulation times with change in computing hardware.
- Both transient performance characteristics and frequency domain analysis should be outlined, underlining whether the algorithm results in an economic scheme or not.
- In addition to searching for newer algorithmic procedures, afford needs to be placed to minimize the drawbacks of the current algorithm.
- If possible, the critical simulation steps should be addressed with the flowchart.
- In a case where control variables to be used during recursion, the variable's initialization and selection procedure should be addressed clearly.

## VIII.  Conclusion

The meta-heuristic algorithms have faced a tremendous phase change in the last few years. Researchers have tried to best observe the tuning in various scenarios by imitating different physical and natural phenomena. Here, we have performed a review investigation on thirty major metaheuristic algorithms for PID gain tuning in an AVR system. It is seen that the newer algorithms are becoming better than the previous ones. Similar research environments and presets are required to perform any comparative analysis among the algorithms correctly. In future investigations, a robust system with Model Predictive Control (MPC) and Linear Quadratic Gaussian (LQG) controllers needs consideration too. Both relative performance and economic feasibility during the implementation of any newer algorithm will help better understand the actual contribution and progress brings forward by that specific algorithm in this control system's horizon.

# References

[1] A. Ula and A. R. Hasan, "Design and implementation of a personal computer based automatic voltage regulator for a synchronous generator," *IEEE Trans. Energy Convers.*, vol. 7, no. 1, pp. 125–131, Mar. 1992, doi: 10.1109/60.124551.

[2] S. Chatterjee and V. Mukherjee, "PID controller for automatic voltage regulator using teaching–learning based optimization technique," *Int. J. Electr. Power Energy Syst.*, vol. 77, pp. 418–429, May 2016, doi: 10.1016/j.ijepes.2015.11.010.

[3] Kiam Heong Ang, G. Chong, and Yun Li, "PID control system analysis, design, and technology," *IEEE Trans. Control Syst. Technol.*, vol. 13, no. 4, pp. 559–576, Jul. 2005, doi: 10.1109/TCST.2005.847331.

[4] Seul Jung and R. C. Dorf, "Analytic PIDA controller design technique for a third order system," in *Proceedings of 35th IEEE Conference on Decision and Control*, Dec. 1996, vol. 3, pp. 2513–2518 vol.3, doi: 10.1109/CDC.1996.573472.

[5] I. Petráš, "Tuning and implementation methods for fractional-order controllers," *Fract. Calc. Appl. Anal.*, vol. 15, no. 2, pp. 282–303, Jun. 2012, doi: 10.2478/s13540-012-0021-4.

[6] P. P. Angelov and D. P. Filev, "An approach to online identification of Takagi-Sugeno fuzzy models," *IEEE Trans. Syst. Man Cybern. Part B Cybern.*, vol. 34, no. 1, pp. 484–498, Feb. 2004, doi: 10.1109/TSMCB.2003.817053.

[7] K. J. Åström and T. Hägglund, "The future of PID control," *Control Eng. Pract.*, vol. 9, no. 11, pp. 1163–1175, Nov. 2001, doi: 10.1016/S0967-0661(01)00062-4.

[8] M. S. Chehadeh and I. Boiko, "Design of rules for in-flight non-parametric tuning of PID controllers for unmanned aerial vehicles," *J. Frankl. Inst.*, vol. 356, no. 1, pp. 474–491, Jan. 2019, doi: 10.1016/j.jfranklin.2018.10.015.

[9] K. J. Åström and T. Hägglund, "Revisiting the Ziegler–Nichols step response method for PID control," *J. Process Control*, vol. 14, no. 6, pp. 635–650, Sep. 2004, doi: 10.1016/j.jprocont.2004.01.002.

[10] A. Rubaai and P. Young, "EKF-Based PI-/PD-Like Fuzzy-Neural-Network Controller for Brushless Drives," *IEEE Trans. Ind. Appl.*, vol. 47, no. 6, pp. 2391–2401, Nov. 2011, doi: 10.1109/TIA.2011.2168799.

[11] F. Olivas, L. Amador-Angulo, J. Perez, C. Caraveo, F. Valdez, and O. Castillo, "Comparative Study of Type-2 Fuzzy Particle Swarm, Bee Colony and Bat Algorithms in Optimization of Fuzzy Controllers," *Algorithms*, vol. 10, no. 3, Art. no. 3, Sep. 2017, doi: 10.3390/a10030101.

[12] V. Mukherjee and S. P. Ghoshal, "Comparison of intelligent fuzzy based AGC coordinated PID controlled and PSS controlled AVR system," *Int. J. Electr. Power Energy Syst.*, vol. 29, no. 9, pp. 679–689, Nov. 2007, doi: 10.1016/j.ijepes.2007.05.002.

[13] A. J. H. Al Gizi, "A particle swarm optimization, fuzzy PID controller with generator automatic voltage regulator," *Soft Comput.*, vol. 23, no. 18, pp. 8839–8853, Sep. 2019, doi: 10.1007/s00500-018-3483-4.

[14] Zwe-Lee Gaing, "A particle swarm optimization approach for optimum design of PID controller in AVR system," *IEEE Trans. Energy Convers.*, vol. 19, no. 2, pp. 384–391, Jun. 2004, doi: 10.1109/TEC.2003.821821.

[15] H. Gozde and M. C. Taplamacioglu, "Comparative performance analysis of artificial bee colony algorithm for automatic voltage regulator (AVR) system," *J. Frankl. Inst.*, vol. 348, no. 8, pp. 1927–1946, Oct. 2011, doi: 10.1016/j.jfranklin.2011.05.012.

[16] D. K. Sambariya, R. Gupta, and R. Prasad, "Design of optimal input–output scaling factors based fuzzy PSS using bat algorithm," *Eng. Sci. Technol. Int. J.*, vol. 19, no. 2, pp. 991–1002, Jun. 2016, doi: 10.1016/j.jestch.2016.01.006.

[17] F. A. G. S. babu and S. B. T. Chiranjeevi, "Implementation of Fractional Order PID Controller for an AVR System Using GA and ACO Optimization Techniques," *IFAC-Pap.*, vol. 49, no. 1, pp. 456–461, Jan. 2016, doi: 10.1016/j.ifacol.2016.03.096.

[18] Z. Bingul and O. Karahan, "A novel performance criterion approach to optimum design of PID controller using cuckoo search algorithm for AVR system," *J. Frankl. Inst.*, vol. 355, no. 13, pp. 5534–5559, Sep. 2018, doi: 10.1016/j.jfranklin.2018.05.056.

[19] S. Panda, B. K. Sahu, and P. K. Mohanty, "Design and performance analysis of PID controller for an automatic voltage regulator system using simplified particle swarm optimization," *J. Frankl. Inst.*, vol. 349, no. 8, pp. 2609–2625, Oct. 2012, doi: 10.1016/j.jfranklin.2012.06.008.

[20] C. Li, H. Li, and P. Kou, "Piecewise function based gravitational search algorithm and its application on parameter identification of AVR system," *Neurocomputing*, vol. 124, pp. 139–148, Jan. 2014, doi: 10.1016/j.neucom.2013.07.018.

[21] U. Güvenç, A. H. Işik, T. Yiğit, and İ. Akkaya, "Performance analysis of biogeography-based optimization for automatic voltage regulator system," *Turk. J. Electr. Eng. Comput. Sci.*, vol. 24, no. 3, pp. 1150–1162, Mar. 2016.

[22] R. V. Rao, V. J. Savsani, and D. P. Vakharia, "Teaching–Learning-Based Optimization: An optimization method for continuous nonlinear large scale problems," *Inf. Sci.*, vol. 183, no. 1, pp. 1–15, Jan. 2012, doi: 10.1016/j.ins.2011.08.006.

[23] P. Yadav, R. Kumar, S. K. Panda, and C. S. Chang, "An Intelligent Tuned Harmony Search algorithm for optimisation," *Inf. Sci.*, vol. 196, pp. 47–72, Aug. 2012, doi: 10.1016/j.ins.2011.12.035.

[24] D. H. Kim, "Hybrid GA–BF based intelligent PID controller tuning for AVR system," *Appl. Soft Comput.*, vol. 11, no. 1, pp. 11–22, Jan. 2011, doi: 10.1016/j.asoc.2009.01.004.

[25] P. K. Mohanty, B. K. Sahu, and S. Panda, "Tuning and Assessment of Proportional–Integral–Derivative Controller for an Automatic Voltage Regulator System Employing Local Unimodal Sampling Algorithm," *Electr. Power Compon. Syst.*, vol. 42, no. 9, pp. 959–969, Jul. 2014, doi: 10.1080/15325008.2014.903546.

[26] B. Hekimoğlu and S. Ekinci, "Grasshopper optimization algorithm for automatic voltage regulator system," in *2018 5th International Conference on Electrical and Electronic Engineering (ICEEE)*, May 2018, pp. 152–156, doi: 10.1109/ICEEE2.2018.8391320.

[27] H. Zhu, L. Li, Y. Zhao, Y. Guo, and Y. Yang, "CAS algorithm-based optimum design of PID controller in AVR system," *Chaos Solitons Fractals*, vol. 42, no. 2, pp. 792–800, Oct. 2009, doi: 10.1016/j.chaos.2009.02.006.

[28] O. Bendjeghaba, "CONTINUOUS FIREFLY ALGORITHM FOR OPTIMAL TUNING OF PID CONTROLLER IN AVR SYSTEM," *J. Electr. Eng.*, vol. 65, no. 1, pp. 44–49, Jan. 2014, doi: 10.2478/jee-2014-0006.

[29] S. Ekinci, B. Hekimoğlu, and S. Kaya, "Tuning of PID Controller for AVR System Using Salp Swarm Algorithm," in *2018 International Conference on Artificial Intelligence and Data Processing (IDAP)*, Sep. 2018, pp. 1–6, doi: 10.1109/IDAP.2018.8620809.

[30] D. K. Sambariya and T. Gupta, "Optimal design of PID controller for an AVR system using monarch butterfly optimization," in *2017 International Conference on Information, Communication, Instrumentation and Control (ICICIC)*, Aug. 2017, pp. 1–6, doi: 10.1109/ICOMICON.2017.8279106.

[31] R. Pradhan, S. K. Majhi, and B. B. Pati, "Design of PID controller for automatic voltage regulator system using Ant Lion Optimizer," *World J. Eng.*, vol. 15, no. 3, pp. 373–387, Jan. 2018, doi: 10.1108/WJE-05-2017-0102.

[32] S. Anbarasi and S. Muralidharan, "Enhancing the Transient Performances and Stability of AVR System with BFOA Tuned PID Controller," *J. Control Eng. Appl. Inform.*, vol. 18, no. 1, Art. no. 1, Mar. 2016.

[33] E. Çelik and R. Durgut, "Performance enhancement of automatic voltage regulator by modified cost function and symbiotic organisms search algorithm," *Eng. Sci. Technol. Int. J.*, vol. 21, no. 5, pp. 1104–1111, Oct. 2018, doi: 10.1016/j.jestch.2018.08.006.

[34] H. M. Hasanien, "Design Optimization of PID Controller in Automatic Voltage Regulator System Using Taguchi Combined Genetic Algorithm Method," *IEEE Syst. J.*, vol. 7, no. 4, pp. 825–831, Dec. 2013, doi: 10.1109/JSYST.2012.2219912.

[35] H. Shayeghi, "Anarchic Society Optimization Based PID Control of an Automatic Voltage Regulator (AVR) System," *Electr. Electron. Eng.*, vol. 2, Aug. 2012, doi: 10.5923/j.eee.20120204.05.

[36] B. K. Sahu, S. Panda, P. K. Mohanty, and N. Mishra, "Robust analysis and design of PID controlled AVR system using Pattern Search algorithm," in *2012 IEEE International Conference on Power Electronics, Drives and Energy Systems (PEDES)*, Dec. 2012, pp. 1–6, doi: 10.1109/PEDES.2012.6484294.

[37] I. Chiha, J. Ghabi, and N. Liouane, "Tuning PID controller with multi-objective differential evolution," in *2012 5th International Symposium on Communications, Control and Signal Processing*, May 2012, pp. 1–4, doi: 10.1109/ISCCSP.2012.6217801.

[38] L. dos Santos Coelho, "Tuning of PID controller for an automatic regulator voltage system using chaotic optimization approach," *Chaos Solitons Fractals*, vol. 39, no. 4, pp. 1504–1514, Feb. 2009, doi: 10.1016/j.chaos.2007.06.018.

[39] E. Çelik, "Incorporation of stochastic fractal search algorithm into efficient design of PID controller for an automatic voltage regulator system," *Neural Comput. Appl.*, vol. 30, no. 6, pp. 1991–2002, Sep. 2018, doi: 10.1007/s00521-017-3335-7.

[40] J. Bhookya and R. K. Jatoth, "Optimal FOPID/PID controller parameters tuning for the AVR system based on sine–cosine-algorithm," *Evol. Intell.*, vol. 12, no. 4, pp. 725–733, Dec. 2019, doi: 10.1007/s12065-019-00290-x.

[41] N. Razmjooy, M. Khalilpour, and M. Ramezani, "A New Meta-Heuristic Optimization Algorithm Inspired by FIFA World Cup Competitions: Theory and Its Application in PID Designing for AVR System," *J. Control Autom. Electr. Syst.*, vol. 27, no. 4, pp. 419–440, Aug. 2016, doi: 10.1007/s40313-016-0242-6.

[42] S. Ekinci and B. Hekimoğlu, "Improved Kidney-Inspired Algorithm Approach for Tuning of PID Controller in AVR System," *IEEE Access*, vol. 7, pp. 39935–39947, 2019, doi: 10.1109/ACCESS.2019.2906980.

[43] E. Köse, "Optimal Control of AVR System With Tree Seed Algorithm-Based PID Controller," *IEEE Access*, vol. 8, pp. 89457–89467, 2020, doi: 10.1109/ACCESS.2020.2993628.

[44] H. Shayeghi, A. Younesi, and Y. Hashemi, "Optimal design of a robust discrete parallel FP+FI+FD controller for the Automatic Voltage Regulator system," *Int. J. Electr. Power Energy Syst.*, vol. 67, pp. 66–75, May 2015, doi: 10.1016/j.ijepes.2014.11.013.

[45] E. Rashedi, H. Nezamabadi-pour, and S. Saryazdi, "GSA: A Gravitational Search Algorithm," *Inf. Sci.*, vol. 179, no. 13, pp. 2232–2248, Jun. 2009, doi: 10.1016/j.ins.2009.03.004.

[46] M. S. Kiran, "TSA: Tree-seed algorithm for continuous optimization," *Expert Syst. Appl.*, vol. 42, no. 19, pp. 6686–6698, Nov. 2015, doi: 10.1016/j.eswa.2015.04.055.

[47] W. K. Ho, K. W. Lim, and W. Xu, "Optimal Gain and Phase Margin Tuning for PID Controllers," *Automatica*, vol. 34, no. 8, pp. 1009–1014, Aug. 1998, doi: 10.1016/S0005-1098(98)00032-6.

[48] M. J. Mahmoodabadi and H. Jahanshahi, "Multi-objective optimized fuzzy-PID controllers for fourth order nonlinear systems," *Eng. Sci. Technol. Int. J.*, vol. 19, no. 2, pp. 1084–1098, Jun. 2016, doi: 10.1016/j.jestch.2016.01.010.

[49] M. Hung, L. Shu, S. Ho, S. Hwang, and S. Ho, "A Novel Intelligent Multiobjective Simulated Annealing Algorithm for Designing Robust PID Controllers," *IEEE Trans. Syst. Man Cybern. - Part Syst. Hum.*, vol. 38, no. 2, pp. 319–330, Mar. 2008, doi: 10.1109/TSMCA.2007.914793.

[50] R. A. Krohling and J. P. Rey, "Design of optimal disturbance rejection PID controllers using genetic algorithms," *IEEE Trans. Evol. Comput.*, vol. 5, no. 1, pp. 78–82, Feb. 2001, doi: 10.1109/4235.910467.

[51] M. A. Sahib and B. S. Ahmed, "A new multiobjective performance criterion used in PID tuning optimization algorithms," *J. Adv. Res.*, vol. 7, no. 1, pp. 125–134, Jan. 2016, doi: 10.1016/j.jare.2015.03.004.

[52] D. Simon, "Biogeography-Based Optimization," *IEEE Trans. Evol. Comput.*, vol. 12, no. 6, pp. 702–713, Dec. 2008, doi: 10.1109/TEVC.2008.919004.

[53] D. Du, D. Simon, and M. Ergezer, "Biogeography-based optimization combined with evolutionary strategy and immigration refusal," in *2009 IEEE International Conference on Systems, Man and Cybernetics*, Oct. 2009, pp. 997–1002, doi: 10.1109/ICSMC.2009.5346055.

[54] S. Duman, N. Yörükeren, and İ. H. Altaş, "Gravitational search algorithm for determining controller parameters in an automatic voltage regulator system," *Turk. J. Electr. Eng. Comput. Sci.*, vol. 24, no. 4, pp. 2387–2400, Apr. 2016.

[55] K. M. Passino, "Biomimicry of bacterial foraging for distributed optimization and control," *IEEE Control Syst. Mag.*, vol. 22, no. 3, pp. 52–67, Jun. 2002, doi: 10.1109/MCS.2002.1004010.

[56] D. E. Goldberg, D. E. Goldberg, G. Goldberg David Edward, and V. A. P. of H. D. E. Goldberg, *Genetic Algorithms in Search, Optimization, and Machine Learning*. Addison-Wesley Publishing Company, 1989.

[57] H. Duan, D. Wang, and X. Yu, "Novel approach to nonlinear PID parameter optimization using ant colony optimization algorithm," *J. Bionic Eng.*, vol. 3, no. 2, pp. 73–78, Jun. 2006, doi: 10.1016/S1672-6529(06)60010-3.

[58] S. Mirjalili, "The Ant Lion Optimizer," *Adv. Eng. Softw.*, vol. 83, pp. 80–98, May 2015, doi: 10.1016/j.advengsoft.2015.01.010.

[59] G. Wang, X. Zhao, and S. Deb, "A Novel Monarch Butterfly Optimization with Greedy Strategy and Self-Adaptive," in *2015 Second International Conference on Soft Computing and Machine Intelligence (ISCMI)*, Nov. 2015, pp. 45–50, doi: 10.1109/ISCMI.2015.19.

[60] X.-S. Yang and S. Deb, "Engineering optimisation by cuckoo search," *Int. J. Math. Model. Numer. Optim.*, vol. 1, no. 4, pp. 330–343, Jan. 2010, doi: 10.1504/IJMMNO.2010.03543.

[61] S. Saremi, S. Mirjalili, and A. Lewis, "Grasshopper Optimisation Algorithm: Theory and application," *Adv. Eng. Softw.*, vol. 105, pp. 30–47, Mar. 2017, doi: 10.1016/j.advengsoft.2017.01.004.

[62] H. Salimi, "Stochastic Fractal Search: A powerful metaheuristic algorithm," *Knowl.-Based Syst.*, vol. 75, pp. 1–18, Feb. 2015, doi: 10.1016/j.knosys.2014.07.025.

[63] M.-Y. Cheng and D. Prayogo, "Symbiotic Organisms Search: A new metaheuristic optimization algorithm," *Comput. Struct.*, vol. 139, pp. 98–112, Jul. 2014, doi: 10.1016/j.compstruc.2014.03.007.

[64] V. F. Yu, A. A. N. P. Redi, C.-L. Yang, E. Ruskartina, and B. Santosa, "Symbiotic organisms search and two solution representations for solving the capacitated vehicle routing problem," *Appl. Soft Comput.*, vol. 52, pp. 657–672, Mar. 2017, doi: 10.1016/j.asoc.2016.10.006.

[65] I. A. Khan, A. S. Alghamdi, T. A. Jumani, A. Alamgir, A. B. Awan, and A. Khidrani, "Salp Swarm Optimization Algorithm-Based Fractional Order PID Controller for Dynamic Response and Stability Enhancement of an Automatic Voltage Regulator System," *Electronics*, vol. 8, no. 12, Art. no. 12, Dec. 2019, doi: 10.3390/electronics8121472.

[66] S. Mirjalili, "SCA: A Sine Cosine Algorithm for solving optimization problems," *Knowl.-Based Syst.*, vol. 96, pp. 120–133, Mar. 2016, doi: 10.1016/j.knosys.2015.12.022.

[67] B. Hekimoğlu, "Sine-cosine algorithm-based optimization for automatic voltage regulator system," *Trans. Inst. Meas. Control*, vol. 41, p. 014233121881145, Nov. 2018, doi: 10.1177/0142331218811453.

[68] N. S. Jaddi and S. Abdullah, "Optimization of neural network using kidney-inspired algorithm with control of filtration rate and chaotic map for real-world rainfall forecasting," *Eng. Appl. Artif. Intell.*, vol. 67, no. C, pp. 246–259, Jan. 2018, doi: 10.1016/j.engappai.2017.09.012.

[69] S. Ekinci, A. Demirören, H. L. Zeynelgil, and S. Kaya, "Design of PID Controller for Automatic Voltage Regulator System through Kidney-inspired Algorithm," *Gazi Üniversitesi Fen Bilim. Derg.*, vol. 7, no. 2, pp. 383–398, Jun. 2019, doi: 10.29109/gujsc.516424.

[70] A. Kumar and G. Shankar, "Priority based optimization of PID controller for automatic voltage regulator system using gravitational search algorithm," in *2015 International Conference on Recent Developments in Control, Automation and Power Engineering (RDCAPE)*, Mar. 2015, pp. 292–297, doi: 10.1109/RDCAPE.2015.7281412.

[71] S. Ekinci, B. Hekimoğlu, and E. Eker, "Optimum Design of PID Controller in AVR System Using Harris Hawks Optimization," in *2019 3rd International Symposium on Multidisciplinary Studies and Innovative Technologies (ISMSIT)*, Oct. 2019, pp. 1–6, doi: 10.1109/ISMSIT.2019.8932941.

[72] S. Ekinci, A. Demirören, H. Zeynelgil, and B. Hekimoğlu, "An opposition-based atom search optimization algorithm for automatic voltage regulator system," 2020, doi: 10.17341/gazimmfd.598576.